\documentclass[a4paper,12pt]{article}

\usepackage{amssymb,amsmath,empheq,epsfig,graphicx}

\usepackage{color}
\newcommand{\At}{{\boldsymbol{A}}}
\newcommand{\Acb}{\tilde{\Ac}}             % new variable A in app.B
\newcommand{\Ac}{\mathcal{A}}            % new variable A in app.B
\newcommand{\ac}{\mathcal{A}}            % reduced action
\newcommand{\ag}{{\alpha_G}}             % transplackian coupling
\newcommand{\ael}{{M_1}}                 % single-exchange amplitude
\newcommand{\amp}{M}                     % amplitude
\newcommand{\ampRid}{\mathfrak{M}}       % reduced amplitude occurring in coherent state
\newcommand{\binomial}[2]{{{#1}\choose{#2}}}  % binomial
\newcommand{\bk}[1]{\langle #1 \rangle}  % bra-ket
\newcommand{\bs}{\boldsymbol}
\newcommand{\bt}{{\boldsymbol{b}}}
\newcommand{\C}{\mathbb{C}}              % complex
\newcommand{\cl}{\mathrm{cl}}            % classico
\newcommand{\deltap}{\delta_{+}}          % delta mass-shell
\newcommand{\dif}{\mathrm{d}}            % finite-dimensional differential
\newcommand{\e}{\varepsilon}             % antisymmetric tensor
\newcommand{\el}{{\mathrm{el}}}          % elastic
\newcommand{\esp}[1]{\mathrm{e}^{#1}}    % esponential
\newcommand{\GW}{\mathrm{GW}}            % gravitatonal wave
\renewcommand{\hom}{\hbar\omega}
\newcommand{\ini}{{\mathrm{in}}}          % incoming
\newcommand{\ket}[1]{|#1\rangle}         % |   >
\newcommand{\lra}{\leftrightarrow}        % scambio
\newcommand{\match}{{\mathrm{matched}}}  % matched
\newcommand{\Num}{\mathcal{N}}           % number density
\newcommand{\om}{\omega}
\newcommand{\omE}{\frac{\hom}{E}}

\newcommand{\ord}[1]{\mathcal{O}\left(#1\right)}
\newcommand{\pol}{\epsilon}              % polarization
\newcommand{\Qt}{{\boldsymbol{Q}}}
\newcommand{\qt}{{\boldsymbol{q}}}
\newcommand{\pt}{{\boldsymbol{p}}}
\newcommand{\R}{\mathbb{R}}              % real
\newcommand{\ram}{\mathrm{R}}
\newcommand{\regge}{\mathrm{Regge}}
\newcommand{\rung}{\mathcal{R}}           % rung
\newcommand{\soft}{\mathrm{soft}}
\newcommand{\tb}{\bar{\tau}}             % saddle point
\newcommand{\tfa}{{\cal M}}              % Fourier transform of amplitude
\newcommand{\Tht}{{\boldsymbol{\Theta}}}
\newcommand{\tht}{{\boldsymbol{\theta}}}
\newcommand{\Usoft}{\mathcal{U}_{\text{soft}}} % imaginary unit
\newcommand{\ui}{\mathrm{i}}             % imaginary unit
\newcommand{\vp}{\vec{p}}                % 3-vector p
\newcommand{\vq}{\vec{q}}                % 3-vector q
\newcommand{\xt}{{\boldsymbol{x}}}
\newcommand{\Z}{\mathbb{Z}}              % integers
\newcommand{\zl}{|z|^2\esp{\ui\lambda\phi_z}}              % z(*)^2 depending on lambda
\newcommand{\Zt}{{\boldsymbol{Z}}}
\newcommand{\zt}{{\boldsymbol{z}}}

\numberwithin{equation}{section}

\title{%\preprint%
{\bf Radiation enhancement and ``temperature'' in the collapse regime of
  gravitational scattering}}

\author{
   Marcello~Ciafaloni
   \footnote{Email: ciafaloni@fi.infn.it}
   \\
   {\sl\small Dipartimento di Fisica, Universit\`a di Firenze}
   \\[1ex]
   and
   \\[1ex]
   Dimitri~Colferai
   \footnote{Email: colferai@fi.infn.it}
   \\
   {\sl\small Dipartimento di Fisica, Universit\`a di Firenze and INFN Firenze}\\
   {\sl\small Via Sansone 1, 50019 Sesto Fiorentino, Italy}\\[5mm]
   \\[5mm]
}
\date{}

\begin{document}
%@@@@@@@@@@@@@@@@@@@@@@@@@@@@@@@@@@@@@@@@@@@@@@@@@@@@@@@@@@@@@@@@@@@@@@@@@@@@@@@

\maketitle

\begin{abstract}
  We generalize the semiclassical treatment of graviton radiation to
  gravitational scattering at very large energies $\sqrt{s}\gg m_P$ and finite
  scattering angles $\Theta_s$, so as to approach the collapse regime of impact
  parameters $b \simeq b_c \sim R\equiv 2G\sqrt{s}$. Our basic tool is the
  extension of the recently proposed, unified form of radiation to the ACV
  reduced-action model and to its resummed-eikonal exchange. By superimposing
  that radiation all-over eikonal scattering, we are able to derive the
  corresponding (unitary) coherent-state operator. The resulting graviton
  spectrum, tuned on the gravitational radius $R$, fully agrees with previous
  calculations for small angles $\Theta_s\ll 1$ but, for sizeable angles
  $\Theta_s(b)\leq \Theta_c = \ord{1}$ acquires an exponential cutoff of the
  large $\om R$ region, due to energy conservation, so as to emit a finite
  fraction of the total energy. In the approach-to-collapse regime of
  $b\to b_c^+$ we find a radiation enhancement due to large tidal forces, so
  that the whole energy is radiated off, with a large multiplicity
  $\bk{N}\sim Gs \gg 1$ and a well-defined frequency cutoff of order $R^{-1}$.
  The latter corresponds to the Hawking temperature for a black hole of mass
  notably smaller than $\sqrt{s}$.
\end{abstract}

%\begin{scriptsize}
%   Draft $ $Revision: 282 2b14beac9e6e $ $ \hspace{20mm}
%   $ $Date: 2016/12/20 22:31:20 $ $
%\end{scriptsize}

\newpage

\tableofcontents

\newpage

%%%%%%%%%%%%%%%%%%%%%%%%%%%%%%%%%%%%%%%%%%%%%%%%%%%%%%%%%%%%%%%%%%%%%%%%%%%%%%%%%
\section{Introduction\label{s:intro}}
%%%%%%%%%%%%%%%%%%%%%%%%%%%%%%%%%%%%%%%%%%%%%%%%%%%%%%%%%%%%%%%%%%%%%%%%%%%%%%%%%

The investigation of transplanckian-energy gravitational scattering performed
since the eighties
\cite{tHooft:1987rb,Muzinich:1987in,ACV87,GrMe87,ACV88,VeVe91,ACV90,ACV93} and
applied to the collapse regime~\cite{ACV07,Marchesini:2008yh,VeWo1,VeWo2} has
been recently revived at both classical~\cite{GrVe14} and quantum
level~\cite{Dvali:2014ila,CCV15} with the purpose of describing the radiation
associated to extreme energies and of gaining a better understanding of a
possibly collapsing system. A bridge between the different approaches
of~\cite{Dvali:2014ila} and~\cite{CCV15} has also been
devised~\cite{Addazi:2016ksu}.

Here we follow essentially the ACV path~\cite{ACV88,ACV90,ACV93,ACV07}, that is
mostly an effective theory based on $s$-channel iteration (eikonal scattering)
and motivated by the smallness of fixed-angle amplitudes in
string-gravity~\cite{GrMe87} and by the high-energy dominance of the spin-2
graviton exchange, at small momentum
transfers~\cite{tHooft:1987rb,Muzinich:1987in,ACV87}. In fact, a key feature of
eikonal scattering is that the large momentum transfers built up at fixed
scattering angle (e.g.\ the Einstein deflection angle $\Theta_E\equiv 2R/b$) ---
$R\equiv 2G\sqrt{s}$ being the gravitational radius --- is due to a large number
$\bk{n}=Gs/\hbar\equiv\ag\gg 1$ of single-hits with very small scattering angle
$\theta_m\sim\hbar/bE$. By following these lines, ACV~\cite{ACV93} proposed an
all-order generalization of the semiclassical approach based on an effective
action~\cite{Li91,VeVe91}, that allows in principle to compute corrections to
the eikonal functions depending on the expansion parameter $R^2/b^2$ (by
neglecting, in string-gravity, the smaller ones
$\ord{l_s^2/b^2}$~\cite{ACV88,Giddings:2007bw} if $l_P < l_s \ll R$).  In its
axisymmetric formulation, the eikonal resummation reduces to a solvable model in
one-dimensional radial space, that was worked out explicitly in~\cite{ACV07}.
Such reduced-action model allows to treat sizeable angles $R/b$, up to a
singularity point $b_c=\sqrt{3\sqrt{3}/2}R$ where a branch point of critical
index $3/2$ occurs in the action, as signal of a possible classical collapse.

The main purpose of the present paper is to extend the radiation treatment
of~\cite{CCV15,CCCV15} to larger angles, by applying it to the ACV resummed
eikonal, in order to achieve a comparable progress at radiation level. We shall
then use it to study the extreme energy regime of a possible classical collapse
$b\to b_c^+\sim R$.

Let us recall that the main qualitative understanding of~\cite{CCV15}, compared
to previous approaches, was to disentangle the role of the gravitational radius
$R$ in the radiation process. In fact, by superimposing the radiation amplitudes
associated to the various eikonal exchanges and by combining the large number
$\bk{n}\sim\ag=ER$ of emitters with the relatively small energy fraction
$\hbar\om/E$, CCV found that the relevant variable becomes $\om R$, which is
thus needed to describe the interference pattern of the whole amplitude
(sec.~\ref{s:gbsa}). In the present paper we follow the same strategy, by
replacing the leading eikonal (single graviton exchange) by the resummed one
(sec.~\ref{s:rm}).

There is, however, an important technical point to be understood. The
single-exchange radiation amplitude was determined in~\cite{CCV15,CCCV15} by
unifying in the $E\gg\hbar\om$ regime the Regge region of large emission angles
with the soft one. Such unifying relationship involves a simple rescaling
$E\to\hbar\om$ of the soft amplitude and is exact for single-graviton exchange.
Here we wish to generalize the soft-based representation so obtained to all
subleading eikonal contributions. No real proof of that statement is available
yet. Nevertheless, we shall argue in sec.~\ref{s:rm} that, starting from the
H-diagram~\cite{ACV90}, the dominant Regge contributions are confined to the
deep fragmentation regions of the incoming particles, thus allowing the
approximate use of the unifying relationship mentioned before and of the ensuing
soft-based representation.

By entering the large angle region, we meet the issue of energy conservation
also~\cite{Giudice:2001ce,Rychkov_pc,Giddings:2004xy,GriVe_pc}. Indeed, the coherent
radiation state obtained by the soft-based formulation treats the fast particles
as sources and thus neglects, in a first instance, conservation constraints. By
introducing them explicitly in sec.~\ref{s:far}, we keep neglecting correlations
that we argue to be small~(sec.~\ref{s:ce}). However, the overall effect of
energy conservation is quite important, in the large-angle region, because it
introduces an exponential cutoff which --- though preserving quantum coherence
--- plays a role similar to the temperature in a statistical ensemble.

The validity of the exponential behaviour and its role in approaching collapse
are carefully discussed in sec.~\ref{s:ecc}. The final outcome is that the whole
energy is radiated off in the approach-to-collapse regime, by fixing the
analogue of the Hawking temperature~\cite{Hawking:1974sw,Hawking:2015qqa} for
our energetic sample of (coherent) radiation.

%%%%%%%%%%%%%%%%%%%%%%%%%%%%%%%%%%%%%%%%%%%%%%%%%%%%%%%%%%%%%%%%%%%%%%%%%%%%%%%%%
\section{Graviton radiation in small-angle transplanckian\\ scattering\label{s:gbsa}}
%%%%%%%%%%%%%%%%%%%%%%%%%%%%%%%%%%%%%%%%%%%%%%%%%%%%%%%%%%%%%%%%%%%%%%%%%%%%%%%%%

The approach to gravitational scattering and radiation advocated in~\cite{CCCV15}
is based on a semiclassical approximation to the $S$-matrix of the form
\begin{equation}\label{Smatrix}
  S \simeq \exp\left\{
    2\ui\hat\delta\Big(\ag,\frac{R}{b},a_\lambda(\vq)\Big)\right\} \;,
\end{equation}
where the eikonal operator $\hat\delta$ is a function of the effective coupling
$\ag\equiv Gs/\hbar$ and of the angular variable $R/b$ (where $R\equiv
2G\sqrt{s}$ is the gravitational radius and $b$ the impact parameter) and a
functional of the graviton step operators $a_\lambda(\vq)$ with helicity
$\lambda$ and momentum $\vq$.

The semiclassical form~\eqref{Smatrix} was argued in~\cite{ACV07} to be valid in
the strong-gravity regime $\ag \gg 1$ with $R \gg b \gg l_s > l_P$, where $l_s$
is the string length and $l_P\equiv\sqrt{\hbar G}$ the Planck length. This means
that we are, to start with, in the transplanckian regime
$\sqrt{s} \gg m_P = \hbar/l_P$ at small scattering angles
$\Theta_s \simeq \Theta_E \equiv 2R/b$, where $\Theta_E$ is the Einstein deflection
angle. Quantum corrections to~\eqref{Smatrix} will involve the parameter
$l_P^2/b^2$ (and $l_s^2/b^2$ if working within string-gravity) and will be
partly considered later on.

The eikonal operator is then obtained by resumming an infinite series of
effective diagrams which include virtual graviton exchanges and real graviton
emissions, as will be shortly reviewed in the following.  In the small-angle and
low-density limit, it is composed by two terms: {\it (i)} a c-number phase shift
$\delta_0$ generated by graviton exchanges between the external particles
undergoing the scattering process; {\it(ii)} a linear superposition of creation
and destruction operators which is responsible of graviton bremsstrahlung and
associated quantum virtual corrections:
\begin{equation}\label{deltahat}
  \hat\delta\Big(\ag,\frac{R}{b},a_\lambda(\vq)\Big) = \delta_0(b) +
  \int\frac{\dif^3 q}{\hbar^3\sqrt{2\om}}\; \sum_{\lambda=\pm 2} \left[
    \ampRid_\lambda(\bt,\vq) a^\dagger_\lambda(\vq) +\text{h.c.}\right]
  + \ord{a_\lambda^2} \;,
\end{equation}
where higher powers of $a_\lambda$ provide high-density corrections.  This
structure, which is valid for large impact parameters, i.e., for small values of
the ratio $R/b \ll 1$, provides a unitary $S$-matrix describing the Einstein
deflection of the scattered particles as well as its associated graviton
radiation and its metric fields~\cite{ACV07,CC14} and time
delays~\cite{Camanho:2014apa}.

Actually, the subject of this paper is to extend the above picture to smaller impact
parameters $b\sim R$ where the gravitational interaction becomes really strong
and a gravitational collapse is expected on classical grounds. We will show
that, decreasing the impact parameter $b$ up to some critical parameter $b_c
\sim R$ of the order of the gravitational radius $R$, the form of the $S$-matrix
maintains the form~(\ref{Smatrix},\ref{deltahat}) with calculable corrections to
both the phase shift $\delta_0$ (sec.~\ref{s:ram}) and to the emission amplitude
$\ampRid$ (sec.~\ref{s:sbr}). We shall then discuss in detail (sec.~\ref{s:far})
what happens in the limit $b\to b_c$ from above.

%===============================================================================
\subsection{Eikonal scattering\label{s:es}}
%===============================================================================

\begin{figure}[t]
  \centering
  \includegraphics[width=0.7\textwidth]{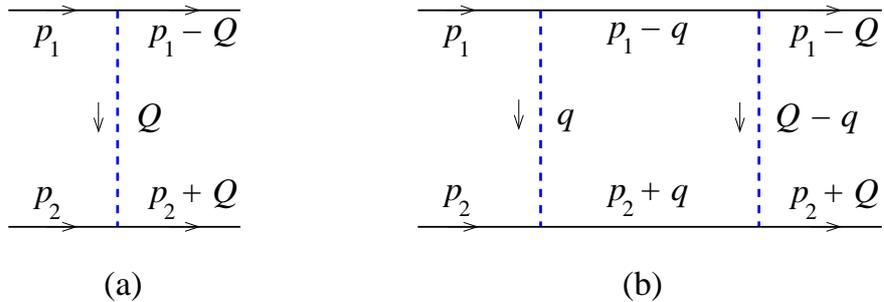}
  \caption{\it One- and two-rung effective ladder diagrams determining the elastic
    $S$-matrix in the eikonal approximation. Solid lines: on-shell external
    particles; dashed lines: eikonal gravitational exchanges.}
  \label{f:ladder}
\end{figure}

ACV~\cite{ACV90} have shown that the leading contributions to the high-energy
elastic scattering amplitude $p_1+p_2\to p'_1+p'_2$ come from the $s$-channel
iteration of soft-graviton exchanges, which can be represented by effective
ladder diagrams as in fig.~\ref{f:ladder}. The generic ladder is built by
iteration (i.e., 4D convolution) of the basic rung
\begin{equation}\label{rung}
 \rung_1(p_1,p_2,Q) = \ui \ael(Q^2,s) \,
 2\pi\deltap\left((p_1-Q)^2\right)2\pi\deltap\left((p_2+Q)^2\right) \;, \quad
 \ael(Q^2,s)\equiv -\frac{8\pi G s^2}{Q^2}
\end{equation}
which embodies the on-shell conditions of the scattered particles and the
Newton-like elastic scattering amplitude $\ael$ in momentum space. The on-shell
conditions and the particular form of $\ael$ make it possible to express the
$n$-rung amplitude as a 2D convolution in the form
\begin{equation}\label{Meln}
  \ui\amp_n(Q^2,s) = \frac{\ui^n}{n!}\int\frac{\dif^2\qt}{(2\pi)^2}\;
  \amp_{n-1}(-(\Qt-\qt)^2,s) \frac{4\pi Gs}{\qt^2}
  =\frac{2s}{n!} \left[\stackrel{n}{\otimes}\frac{\ui 4\pi Gs}{\Qt^2}\right]\;,
\end{equation}
where boldface variables denote 2D euclidean transverse components.
By Fourier transforming from transverse momentum $\Qt$ to impact
parameter $\bt$, the full eikonal scattering amplitude can be diagonalized and
exponentiated
\begin{equation}\label{Ssum}
  \frac{\ui}{2s}\amp(Q^2,s) \equiv \frac{\ui}{2s}\sum_{n=0}^\infty\amp_n(Q^2,s)
  = \int\dif^2\bt\;\esp{\ui\Qt\cdot\bt} \esp{\ui 2\delta_0(b,s)} \;.
\end{equation}
in terms of the eikonal phase-shift $\delta_0(b,s)$
defined as the Fourier transform of the single-exchange amplitude
\begin{equation}\label{tdfM}
 2\delta_0(b,s) = \int\frac{\dif^2\Qt}{(2\pi)^2}\;
 \esp{-\ui\Qt\cdot\bt} \frac{4\pi Gs}{\Qt^2} \Theta(\Qt^2-Q_0^2)
 = 2 Gs\ln\frac{L}{b} + \ord{\frac{b}{L}}^2 \;,
\end{equation}
where $b\equiv|\bt|$ and $L\equiv 2\esp{-\gamma_E}/Q_0$ is a factorized --- and
thus irrelevant --- infrared cutoff needed to regularize the ``Coulomb''
divergence typical of long-range interactions.

In order to go beyond the leading eikonal approximation, one has to consider
other diagrams providing corrections of relative order $(R/b)^2$ to elastic
scattering and also inelastic processes (graviton bremsstrahlung). The former
will be dealt with in sec.~\ref{s:rm}; in the following of this section we shall
review graviton bremsstrahlung as derived in~\cite{CCCV15}.

%===============================================================================
\subsection{Unified emission amplitude from single graviton exchange\label{s:uea}}
%===============================================================================

In this subsection we review the derivation of the unified emission amplitude
for the basic process $2\to 2 + \text{graviton}$. ``Unified'' means that such
amplitude is accurate for both large (Regge region) and small (collinear region)
graviton emission angles.

\begin{figure}[ht]
  \centering
  \includegraphics[width=0.7\linewidth]{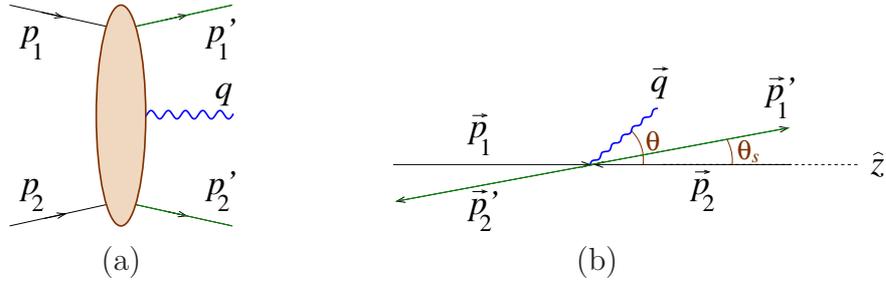}\\
  \hspace{0.18\linewidth}(a) \hspace{0.33\linewidth} (b)\hspace{0.32\linewidth}\null \\
  \caption{\it {\rm (a)} Diagram representing graviton emission (wavy line) in
    transplanckian scattering of two sources (straight lines). {\rm (b)}
    kinematics in the spatial momentum space.}
    \label{f:treeEmission}
\end{figure}

Consider the basic emission process $p_1+p_2 \to p'_1 + p'_2 +q$ at tree level
(fig.~\ref{f:treeEmission}) of a graviton of momentum
$q^\mu:\qt=\hbar\om\tht$ and helicity $\lambda$, assuming a relatively soft
emission energy $\hbar\om\ll E$. Note that this restriction still allows for
a huge graviton phase space, corresponding to classical frequencies potentially
much larger than the characteristic scale $R^{-1}$, due to the large
gravitational charge $\alpha_G \equiv G s / \hbar \gg 1$.

We denote with $\qt_s$ the single-hit transverse momentum exchanged between
particles 1 and 2, and with $\tht_s=|\tht_s|(\cos\phi_s,\sin\phi_s)=\qt_s/E$
the corresponding 2D scattering angle (including azimuth).
For not too large emission angles $|\tht| \ll (E/\hom)|\tht_s|$, corresponding to
$|\qt|\ll|\qt_s|$, Weinberg's theorem expresses the emission amplitude as the
product of the elastic amplitude $\ael$ and of the external-line insertion
factor $J_W^{(\lambda)}\equiv J_W^{\mu\nu}\pol^{(\lambda)*}_{\mu\nu}$, where
$\pol^{(\lambda)}_{\mu\nu}$ is the polarization tensor of the emitted graviton
(see~\cite{CCCV15} for details)
and $J_W^{\mu\nu}$ is the Weinberg current~\cite{We65} [$\eta_i=+1(-1)$ for
incoming (outgoing) lines]
\begin{equation}\label{Jw}
  J_W^{\mu\nu} = \kappa \sum_i \eta_i \frac{p_i^\mu p_i^\nu}{p_i\cdot q}
 = \kappa\left(\frac{p_1^\mu p_1^\nu}{p_1\cdot q}-\frac{p'_1{}^\mu p'_1{}^\nu}{p'_1\cdot q}
 + \frac{p_2^\mu p_2^\nu}{p_2\cdot q}-\frac{p'_2{}^\mu p'_2{}^\nu}{p'_2\cdot q}\right) \;.
\end{equation}
By referring, for definiteness, to the forward hemisphere and restricting
ourselves to the forward region $|\tht|, |\tht_s| \ll 1$ one obtains the
following explicit result~\cite{CCCV15} in the c.m.\ frame with $\pt_1 = 0$:
\begin{equation}\label{Jw+-}
  J_{W}^{(\lambda)}(q^3>0;\tht,\tht_s) = \kappa\frac{E}{\hom} \;
  \left(\esp{\ui\lambda(\phi_\tht-\phi_{\tht-\tht_s})} - 1\right) \; ,
\end{equation}
leading to a factorized soft emission amplitude
\begin{equation}\label{Msoft}
  \amp_{\soft}(\tht_s, E,\tht,\om)
  = \amp_{\el}(E,\Qt) J_W^{(\lambda)}\Big(\frac{E}{\hom},\tht,\tht_s\Big)
  = \kappa^3 s^2 \frac{1}{E \hom\tht_s^2}
  \left(\esp{\ui\lambda(\phi_\tht-\phi_{\tht-\tht_s})} - 1\right) \;,
\end{equation}
The simple expression~\eqref{Jw+-} shows a $1/\om$ dependence, but no
singularities at either $\tht=0$ or $\tht=\tht_s$ as we might have expected from
the $p_i\cdot q$ denominators occurring in~\eqref{Jw}. This is due to the
helicity conservation zeros in the physical projections of the tensor numerators
in~\eqref{Jw}.

The soft amplitude in impact parameter space is readily obtained by Fourier
transforming with respect to $\Qt=\qt_s=E\tht_s$ and reads
\begin{align}
  \tfa_\soft^{(\lambda)}(\bt, E,\tht,\om)
 &\equiv
  \frac1{(2\pi)^{3/2}}\frac1{4s}\int\frac{\dif^2 \qt_s}{(2\pi)^2}\;
  \esp{\ui\qt_s\cdot\bt}\amp_\soft(\qt_s) \nonumber \\
  &= \sqrt{\alpha_G}\frac{R}{\pi} \int\frac{\dif^2\tht_s}{2\pi\tht_s^2}
    \esp{\ui \frac{E}{\hbar}\bt\cdot\tht_s} \frac{E}{\hom}
    \frac{1}{2}
  \left(\esp{\ui\lambda(\phi_{\tht}-\phi_{\tht-\tht_s})} -1\right)\label{Mbsoft} \\
  &= \sqrt{\alpha_G}\frac{R}{\pi} \esp{\ui\lambda\phi_{\tht}} \int
   \frac{\dif^2 \zt}{2\pi\zl}
   \esp{\ui b\om\zt\cdot\tht}
   \frac{E}{\hom} \log\left|\hat{\bt}-\frac{\hom}{E}\zt\right| \;, \nonumber
\end{align}
where in the last line we have used an integral representation which will be
very useful in the sequel.

For large emission angles $|\tht|\gtrsim(E/\hom)|\tht_s|$ such that
$|\qt|\gtrsim|\qt_s|$, graviton emission from internal insertions are no longer
negligible, and Weinberg's formula cannot be applied. However, this region of
phase space is a subset of the so called Regge region, characterized by emission
angles $|\tht|\gg|\tht_s|$. In the Regge limit, the emission amplitude has a
different factorized structure and a different emission current: the Lipatov's
current $J_L^{\mu\nu}$~\cite{Li82}.  Furthermore, one has to distinguish two
transferred momenta $\qt_{1(2)} \equiv \pt_{1(2)} - \pt_{1(2)}'$ such that
$\qt = \qt_1 + \qt_2$.  In the c.m. frame with zero incidence angle
($\pt_1 = E \Tht_i = 0$) and in the forward region $|\tht|,|\tht_s| \ll 1$
(where we identify $\qt_s=\qt_2$), the helicity amplitude takes the
form~\cite{CCCV15}
\begin{align}\label{Mregge}
  \amp_\regge^{(\lambda)}(\tht_s, E,\tht,\om)
 &= \frac{\kappa^2 s^2}{|\qt_1|^2 |\qt_2|^2} J_L^{\mu\nu}
  \pol_{\mu\nu}^{(\lambda)*}  = \kappa^3 s^2
  \frac{1-\esp{\ui\lambda(\phi_{\qt_2}-\phi_{\qt - \qt_2})}}{\qt^2} \;, \\
 \tfa_\regge^{(\lambda)}(\bt,E,\tht,\om)
  &= \sqrt{\alpha_G}\frac{R}{\pi} \esp{\ui\lambda\phi_{\tht}} \int
   \frac{\dif^2 \zt}{2\pi\zl}
   \esp{\ui b\om\zt\cdot\tht} \left(
   -\hat{\bt}\cdot\zt-\log|\hat{\bt}-\zt|\right)\;. \label{Mbregge}
\end{align}
It is not difficult to verify that the soft and Regge amplitudes~\eqref{Msoft},
\eqref{Mregge} agree in the overlapping region of validity
$|\tht_s| \ll \tht \ll (E/\hom)|\tht_s|$. By exploiting the above
expressions, we obtained a unifying amplitude that accurately describes both
regimes and that can be written in terms of the soft amplitude only:
\begin{align}
 \frac{\tfa_\match^{(\lambda)}}{\sqrt{\alpha_G}\frac{R}{\pi} \esp{\ui\lambda\phi_{\tht}}}
 &= \int \frac{\dif^2 \zt}{2\pi\zl}
  \esp{\ui b\om\zt\cdot\tht} \left(\frac{E}{\hom} \log\left| \hat{\bt}
      -\frac{\hom}{E}\zt \right| - \log\left| \hat{\bt} - \zt \right| \right)
  \label{Mmatch}
  &= \soft\big|_E - \soft\big|_{\hom}
\end{align}

The result~\eqref{Mmatch} is expressed in terms of the ($\om$-dependent)
``soft'' field%
\footnote{Notation: the 2D vectors $\Qt,\;\qt,\;\tht,\;\zt,$ etc.\ are
  denoted with boldface characters; their complex representation, e.g.,
  $z=z_1+\ui z_2=|z|\esp{\ui\phi_z}$ is denoted with italic fonts. Note however
  that $b=|\bt|$ is a real quantity.}
\begin{equation}\label{PhiRdef}
  h_s^{(\lambda)}(\om,z) \equiv \frac{1}{\pi^2\zl}
  \left( \frac{E}{\hom} \log\left| \hat{\bt} -\frac{\hom}{E}\zt \right|
    - \log\left| \hat{\bt} - \zt \right| \right)
  \equiv - \frac{\Phi(\om,\zt)}{\pi^2\zl}
\end{equation}
in which the function $\Phi$ turns out to be useful for the treatment of
rescattering too (sec.~\ref{s:eer}). Furthermore, for relatively large
angles ($|\tht| \gg \theta_m \equiv \hbar/(E b)$), eq.~\eqref{Mmatch} involves
values of $\hom|z|/E\lesssim\theta_m/|\tht|$ which are uniformly small,
and the expressions~\eqref{PhiRdef} can be replaced by their $\om\to 0$
limits
\begin{equation}\label{Phidef}
  h_s^{(\lambda)}(z) = -\frac{\Phi_\cl(\zt)}{\pi^2\zl} \;,
  \qquad\Phi_\cl(\zt) \equiv \lim_{\om\to0}\Phi
  = \hat{\bt}\cdot\zt + \log\left|\hat{\bt}-\zt\right| \;,
\end{equation}
which is the field occurring in the Regge amplitude~\eqref{Mbregge}; the
modulating function $\Phi_\cl$ appears also in the classical analysis of
radiation~\cite{GrVe14}.

The last aspect we have to take into account in order to determine the general
$2\to 3$ high-energy emission amplitude at lowest order, is to consider the case
of incoming particles with generic direction of momenta. Since we always work in
the c.m.\ frame, we parametrize $\vp_1=E(\Tht_i,\sqrt{1-\Tht_i^2})$, where
$\Tht_i=|\Tht_i|(\cos\phi_i,\sin\phi_i)$ is a 2D vector that describes both
polar and azimuthal angles of the incoming particles. In~\cite{CCCV15} we proved
the transformation formula for the generic helicity amplitude
\begin{equation}\label{transForm}
 \tfa^{(\lambda)}(\bt,E,\tht,\om;\Tht_i) =
 \esp{\ui\lambda(\phi_\tht-\phi_{\tht-\Tht_i})}
 \tfa^{(\lambda)}(\bt,E,\tht-\Tht_i,\om;\bs{0}) \;.
\end{equation}
By applying eq.~\eqref{transForm} to the matched amplitude~\eqref{Mmatch}, one
immediately finds
\begin{equation}\label{Mmi}
  \tfa_\match(\bt,E,\tht,\om;\Tht_i)
  = \sqrt{\alpha_G}\frac{R}{2} \esp{\ui\lambda\phi_{\tht}} \int
  \dif^2 \zt \; \esp{\ui b\om\zt\cdot(\tht-\Tht_i)} h_s^{(\lambda)}(\om,z)
\end{equation}
and the whole $\Tht_i$ dependence amounts to a shift in the exponential factor.

%===============================================================================
\subsection{Eikonal emission and rescattering\label{s:eer}}
%===============================================================================

\begin{figure}[ht]
  \centering
  \includegraphics[width=0.66\linewidth]{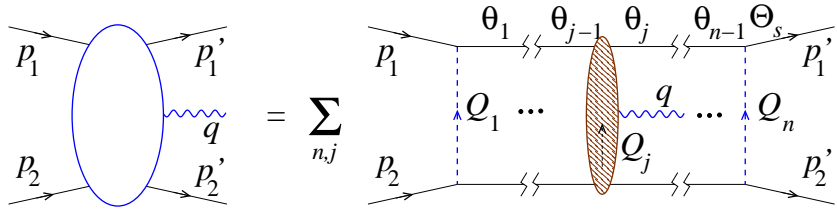}
  \caption{\it Graviton emission from the eikonal ladder. The $n$-rung diagram
    with the emission from the $j$-th exchange is denoted by $\tfa^{[n,j]}$ in
    the text.}
  \label{f:eikEmission}
\end{figure}

The physics of transplanckian scattering is captured, at leading level, by the
resummation of eikonal diagrams, as illustrated in sec.~\ref{s:es}. In order to
compute the associated graviton radiation, it is therefore mandatory to consider
graviton emission from all ladder diagrams, as depicted in
fig.~\ref{f:eikEmission}.

As we shew in ref.~\cite{CCCV15}, the crucial fact is that \emph{all} internal
lines insertions --- for fast particles and exchanged particles alike --- can be
accounted for by calculating $n$ diagrams for the eikonal contribution with $n$
exchanged gravitons, where the \emph{matched} amplitude~\eqref{Mmi} is
inserted in turn in correspondence to the $j$-th exchanged graviton
(fig.~\ref{f:eikEmission}), adjusting for the local incidence angle
$\Tht_i=\tht_{j-1}$.

The ladder-like structure of such amplitude in momentum space is a convolution
in the $\Qt_j$ variables, with $\Qt_1+\cdots+\Qt_n=\Qt$. Thus in impact
parameter space the amplitude is obtained as a product of $j-1$ elastic
amplitudes, the emission amplitude from the $j$-th leg, and $n-j$ elastic
amplitudes, whose upper particle, by energy conservation, has reduced energy
$E\to E-\hom$.

Let us express the elastic amplitude in terms of the dimensionless function
$\Delta(\bt)$ such that
\begin{equation}\label{elAmp}
  \tfa_\el(\bt,E) = 2\delta(\bt,s=4 E E') \equiv 2R \frac{E}{\hbar}\Delta(\bt)
  \;, \qquad \Delta_0(\bt) = \log\left(\frac{L}{b}\right) \;,
\end{equation}
so as to explicitly show the linear proportionality of the amplitude on the
upper (jet 1) particle energy $E$ (which varies after graviton emission).  The
energy $E'$ of the lower particle (jet 2) stays unchanged and its dependence has
been absorbed in the constant $R=4GE'$.

The $n$-rung amplitude for emission of a graviton with momentum $q$ from the
$j$-th exchange of the ladder can then be expressed by the $z$-representation
\begin{align}
  \ui\tfa^{[n,j]}_{\lambda}(\bt,\om,\tht) &= \esp{\ui\lambda\phi_\tht}
  \sqrt{\alpha_G}\frac{R}{2} \times \nonumber \\
  &\quad \frac{\ui^n}{n!}\int\dif^2\zt \;\esp{\ui b\om\tht\cdot\zt}
  \left[\tfa_\el(\bt-\omE b\zt,E)\right]^{j-1} h_s^{(\lambda)}(\om,z)
  \left[\tfa_\el(\bt,E-\hom)\right]^{n-j} \;. \label{zRepAmp}
\end{align}
Note the effect of the incidence angle
$\Tht_i=\tht_{j-1}=(\Qt_1+\cdots+\Qt_{j-1})/E$ in the exponent of
eq.~\eqref{Mmi} which, after Fourier transform, has shifted the impact
parameters of the first elastic amplitudes by the amount $-\omE b\zt$. In
addition, as already mentioned, the energy of the upper particle after the
emission, has the reduced value $E-\hom$, and this modifies the second argument
of the elastic amplitudes after the emission.

\begin{figure}[ht]
  \centering
  \includegraphics[width=0.9\linewidth]{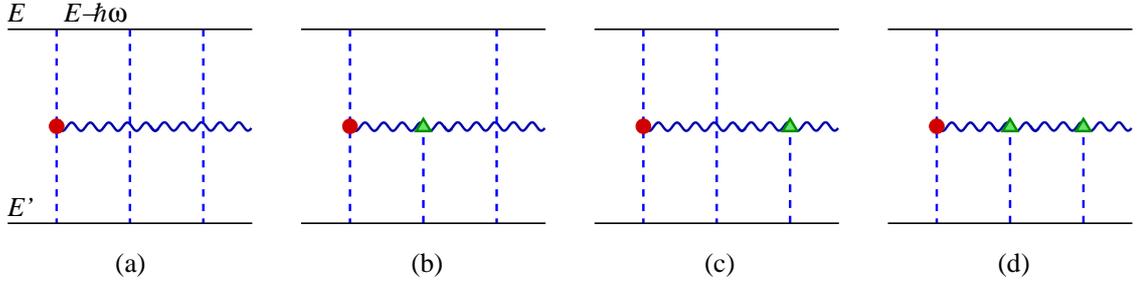}
  \caption{\it Rescattering contributions {\rm(b,c,d)} to eikonal graviton
    emission {\rm (a)}.}
  \label{f:emissionRescatter2}
\end{figure}

Before summing all ladder diagrams, we take into account the rescattering of the
emitted graviton with the external particles $p_j:j=1,2$. This interaction is
proportional to $G(p_j+q)^2$, and is dominated by the exchange of gravitons
between $q$ and $p_2$, since $(p_1+q)^2 \sim \hom E |\tht-\tht_j|^2 \ll
(p_2+q)^2 \sim\hom E$ in the region of forward emission that we are considering.
In practice, we add to the rightmost factor
$\left[\tfa_\el(\bt,E-\hom)\right]^{n-j}$ in eq.~\eqref{zRepAmp} (represented by
the ladder of fig.~\ref{f:emissionRescatter2}.a) the contributions coming from
rescattering diagrams where graviton exchanges between $p_1$ and $p_2$ are
replaced by exchanges between $q$ and $p_2$ in all possible ways
(fig.~\ref{f:emissionRescatter2}.b,c,d).  Since the ordering among eikonal
exchanges and rescattering factors is irrelevant, the inclusion of such
additional contributions amounts to the replacement ($N=n-j$)
\begin{align}
  \left[\tfa_\el(\bt,E-\hom)\right]^N &\to \sum_{r=0}^N
  \binomial{N}{r} \left[\tfa_\el(\bt,E-\hom)\right]^{N-r}
  \left[\tfa_\el(\bt-\xt,\hom)\right]^r \nonumber \\
  &= \left[\tfa_\el(\bt,E-\hom)+\tfa_\el(\bt-\xt,\hom)\right]^N \nonumber \\
  &= \{2R [(E/\hbar-\om)\Delta(\bt)+\om\Delta(\bt-\xt)]\}^N \label{resc}
\end{align}
where we took into account that in diagrams with $N$ exchanged gravitons there
are $\binomial{N}{r}$ distinct diagrams with $r$ rescattering gravitons, and
that in each rescattering factor the energy of the upper particle (i.e., the
emitted graviton) is $\hom$. Furthermore, we took into account that the
transverse position of the emitted graviton with respect to the lower particle
(i.e., $p_2$) is $\bt-\xt$, where $\xt = b\zt$ is the variable conjugated to
$\qt=\om\tht$ [cf.~eq.~\eqref{zRepAmp}], hence to be interpreted as the
transverse position of the emitted graviton with respect to $p_1$.

Substituting the expression of eq.~\eqref{resc} into eq.~\eqref{zRepAmp}, we can
perform the sum over $n$ and $j$ of all diagrams with the aid of the formula
\begin{equation}\label{sum1}
  \sum_{n=1}^\infty \frac1{n!}\sum_{j=1}^n A^{j-1} B^{n-j}
  =\sum_{n=0}^\infty \frac1{n!}\frac{A^n- B^n}{A-B} 
  = \frac{\esp{A}-\esp{B}}{A-B} \;.
\end{equation}
It is also convenient to express the $A$ and $B$ quantities as the elastic
amplitude~\eqref{elAmp} plus a quantum correction $\Phi_{A,B}$ as follows:
\begin{subequations}\label{phiAB}
  \begin{align}
    A &\equiv \ui\tfa_\el(\bt-\omE \xt,E)
    = 2\ui\ag\left[\Delta(\bt)+\omE\Phi_A(\xt)\right] \nonumber \\
    B &\equiv \ui\left[\tfa_\el(\bt,E-\hom)+\tfa_\el(\bt-\xt,\hom)\right]
    = 2\ui\ag\left[\Delta(\bt)+\omE\Phi_B(\xt)\right] \nonumber \\
    \Phi_A(\xt) &\equiv \frac{E}{\hom}[\Delta(\bt-\omE\xt)-\Delta(\bt)] =
    -\Delta'(\bt)\cdot\xt + \ord{\omE} \label{phiA} \\
    \Phi_B(\xt) &\equiv \Delta(\bt-\xt)-\Delta(\bt)] =
    \Phi_A(\xt)|_{E\to\hom}\;. \label{phiB}
  \end{align}
\end{subequations}
We note that the denominator in eq.~\eqref{sum1} is proportional to the $\Phi$
function defined in eq.~\eqref{PhiRdef}
\begin{equation}\label{denom1}
  A-B = 2\ui\ag\omE\left[\Phi_A-\Phi_B\right] = 2\ui\om R\Phi
\end{equation}
and is therefore intimately related to the soft field $h_s$.

From the technical point of view, such relation provides the cancellation
between $h_s$ in eq.~\eqref{zRepAmp} and the mentioned denominator $A-B$
of~\eqref{sum1}, to yield finally the one-graviton emission amplitude
\begin{align}
  \ui\tfa_{\lambda}(\bt;\om,\tht) &= \esp{\ui\ag 2\Delta(b)}
   \ampRid_{\lambda}(\bt;\om,\tht) \nonumber \\
  \frac{\ampRid_{\lambda}(\bt;\om,\tht)}{\esp{\ui\lambda\phi_\tht}}
  &\equiv \sqrt{\alpha_G}\frac{R}{\pi} \int\frac{\dif^2\zt}{2\pi\zl} \;
  \esp{\ui b\om\tht\cdot\zt} \esp{2\ui\om R\Phi_A}
  \frac{\esp{-2\ui\om R \Phi}-1}{2\ui\om R} \;,\label{ampRid2}
\end{align}
which reduces to the classical expression (4.11) of~\cite{GrVe14} in the limit
$\hom/E\to 0$, $\lambda=-2$ and $\Delta=\Delta_0$, since $\Phi\to\Phi_\cl$
[cf.~eq.~\eqref{Phidef}] and
$2R\Phi_A \to 2R\hat{\bt}\cdot\zt=-b\Tht_s\cdot\zt$.

From the conceptual point of view, the identity~\eqref{denom1} is surprising
because it relates the exponents (which describe elastic plus rescattering
exchanges) to the soft field $\sim\Phi$ (which describes graviton
emission). The explanation lies in the derivation~\cite{CCCV15} of the
soft-based representation~\eqref{Mmatch} whose form
\begin{equation}\label{softRegge}
  \tfa_\match =
  \left.\soft\right|_E -\left.\soft\right|_{\hom}
  \simeq \left.\regge\right|_E
\end{equation}
has the alternative interpretations of external plus internal insertions in the
soft-emission language and of elastic plus rescattering ones in the Regge
language.

We shall base on that representation the generalized emission amplitude
including subleading corrections, that will be investigated in
sec.~\ref{s:rm} [eqs.~\eqref{Mres} and \eqref{hsRAM}].

%===============================================================================
\subsection{Multi-graviton emission and linear coherent state\label{s:mge}}
%===============================================================================

\begin{figure}[ht]
  \centering
  \includegraphics[width=0.6\linewidth]{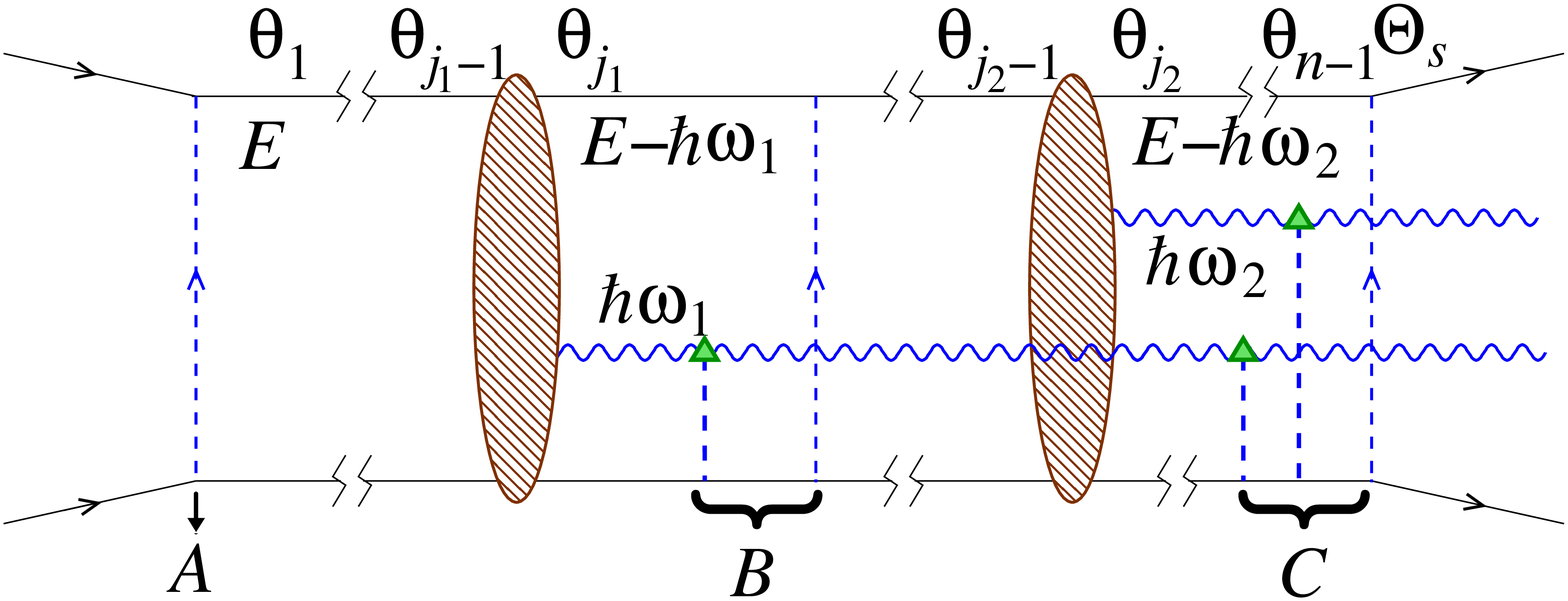}
  \caption{\it Double graviton emission from eikonal ladder. $A$ denotes eikonal
  exchanges before graviton emissions, $B$ eikonal exchanges and rescattering
  of graviton 1, $C$ includes also rescattering of graviton 2.}
  \label{f:eikEmission2g}
\end{figure}

In order to compute the multi-graviton emission amplitude from eikonal ladder
diagrams, let us start from the two-graviton emission process.  We exploit again
the $\bt$-space factorization formula of Regge-amplitudes.  Referring to
fig.~\ref{f:eikEmission2g}, if graviton 1 is emitted first from the $j_1$-th
rung and then graviton 2 from the $j_2$-th rung ($j_1<j_2$) of an $n$-rung
ladder, the corresponding amplitude reads
\begin{align}
  \ui\tfa^{[n,(j_1<j_2)]}(1,2) &= \frac{\ui^n}{2!n!}
    \esp{\ui(\lambda_1\phi_{\tht_1}+\lambda_2\phi_{\tht_2})}
    \left(\sqrt{\ag}\frac{R}{2}\right)^2
    \int\dif^2\xt_1\dif^2\xt_2 \;
    \esp{\ui(\om_1\tht_1\cdot\xt_1+\om_2\tht_2\cdot\xt_2)} \nonumber \\
   &\quad\times A^{j_1-1} h_s(\bt,\xt_1,\om_1) B^{j_2-j_1-1}
    h_s(\bt,\xt_2,\om_2) C^{n-j_2}\label{1<2}
\end{align}
where, as before, the fields $h_s$ describe real graviton production, while
graviton exchanges and rescattering are encoded by the quantities
\begin{align}
  A &= \ui\tfa_\el(\bt-\frac{\hom_1\xt_1+\hom_2\xt_2}{E},E) \nonumber \\
  B &= \ui\tfa_\el(\bt-\frac{\hom_2}{E}\xt_2,E-\hom_1)+\ui\tfa_\el(\bt-\xt_1,\hom_1)
  \nonumber \\
  C &= \ui\tfa_\el(\bt,E-\hom_1-\hom_2)+\ui\tfa_\el(\bt-\xt_1,\hom_1)
  +\ui\tfa_\el(\bt-\xt_2,\hom_2) \;.
\end{align}
The quantity $A$ denotes the eikonal exchanges before any graviton emission, and
is given by the elastic amplitude $\tfa_\el$ with a shift in its first argument
due to the effect of the incidence-angles $\Tht_{j_1-1}, \Tht_{j_2-1}$ of both
gravitons, an effect that propagates backwards in the ladder, as explained in
the previous section.

The quantity $B$ describes the interactions occurring in the middle of the
ladder, i.e., after the emission of graviton 1 and before that of graviton 2.
It consists in the sum of two terms: the first one describes the eikonal
exchanges between $p_1$ and $p_2$, and includes both the effect of the incidence
angle $\Tht_{j_2-1}$ (shift in the first argument) and also the reduced
gravitational coupling $E\to E-\hom_1$ in the upper vertices due to energy
conservation. The second term describes rescattering of graviton 1 with $p_2$.

Finally, the first of the three terms building $C$ represents the eikonal
exchanges between $p_1$ and $p_2$ with reduced coupling $E\to
E-\hbar(\om_1+\om_2)$ in the upper vertices, while the other two terms take into
account rescattering corrections of both gravitons.

The sum over all such ladder diagrams amounts to
\begin{align}
  \mathfrak{S}(1,2) &\equiv \sum_{n=2}^\infty \frac1{n!}\sum_{j_1=1}^{n-1}\sum_{j_2=j_1+1}^n
  A^{j_1-1}B^{j_2-j_1-1}C^{n-j_2} \nonumber \\
  &= \frac{\esp{A}}{(A-B)(A-C)} + \frac{\esp{B}}{(B-A)(B-C)} +
  \frac{\esp{C}}{(C-A)(C-B)} \;. \label{sum2g}
\end{align}
By swapping the graviton indices $1\lra 2$ one immediately obtains the symmetric
contribution with graviton 2 emitted ``before'' graviton 1.

Now, the sum of these two contributions doesn't factorize exactly in two
independent factors. It would if $A-B=[B-C]_{1\lra 2}$, but this is not the
case. However, $A-B=[B-C]_{1\lra 2}+\ord{A\hbar^2\om_i^2/E^2}$, therefore
factorization can be recovered by neglecting contributions of relative order
$\ord{\hbar^2\om_i^2/E^2}$.  In fact, thanks to eqs.~\eqref{phiAB}, we have
\begin{align}
  A &= 2\ui\ag\Delta(\bt-\frac{\hom_1\xt_1+\hom_2\xt_2}{E})
  = 2\ui\ag\left[\Delta(\bt)-\Delta'(b)\cdot
    \frac{\hom_1\xt_1+\hom_2\xt_2}{E}+\ord{\frac{\hbar^2\om_i^2}{E^2}}\right] \nonumber\\
  &= 2\ui\left\{\ag\Delta(\bt)+\om_1 R\Phi_A(\xt_1)
    +\om_2 R\Phi_A(\xt_2)+\ord{G\hbar^2\om_i^2}\right\} \nonumber \\
  B &= 2\ui\ag\left[\left(1-\frac{\hom_1\xt_1}{E}\right)\Delta(\bt-\frac{\hom_2\xt_2}{E})
  +\frac{\hom_1\xt_1}{E}\Delta(\bt-\xt_1) \right] \nonumber \\
  &= 2\ui\left\{\ag\Delta(\bt)+\om_2 R\Phi_A(\xt_2)
    +\om_1 R\Phi_B(\xt_1)+\ord{G\hbar^2\om_i^2}\right\} \nonumber \\
  C &= \ui2\ag\left[\left(1-\frac{\hom_1\xt_1}{E}-\frac{\hom_2\xt_2}{E}\right)\Delta(\bt)
    +\frac{\hom_1\xt_1}{E}\Delta(\bt-\xt_1) +\frac{\hom_1\xt_2}{E}\Delta(\bt-\xt_2)
  \right] \nonumber \\
  &= 2\ui\left\{\ag\Delta(\bt)+\om_1 R\Phi_B(\xt_1)
    +\om_2 R\Phi_B(\xt_2)\right\} \label{ABCexpn}
\end{align}
By noting that the elastic amplitude $\esp{2\ui\ag\Delta(\bt)}$ is a common
factor in all exponentials, we can approximate the infinite sum~\eqref{sum2g}
in the form
\begin{equation}\label{sum12}
  \mathfrak{S}(1,2) \simeq \esp{2\ui\ag\Delta}\left[
    \frac{\esp{\varphi_{A1}+\varphi_{A2}}}{\varphi_1(\varphi_1+\varphi_2)}
    - \frac{\esp{\varphi_{B1}+\varphi_{A2}}}{\varphi_1 \varphi_2}
    + \frac{\esp{\varphi_{B1}+\varphi_{B2}}}{(\varphi_1+\varphi_2)\varphi_2}
  \right] \;,
\end{equation}
where we used the shortcuts $\varphi_{A1}\equiv 2\ui\om_1 R\Phi_A(\xt_1)$ and
analogous ones.

At this point, it is straightforward to check that
\begin{equation}\label{sum21}
  \mathfrak{S}(1,2)+\mathfrak{S}(2,1) \simeq \esp{2\ui\ag\Delta}
  \frac{\esp{\varphi_{A1}}(\esp{-\varphi_1}-1)}{\varphi_1}
  \frac{\esp{\varphi_{A2}}(\esp{-\varphi_2}-1)}{\varphi_2}
\end{equation}
and to obtain the two-graviton emission amplitude in factorized form:
\begin{equation}\label{amp2g}
  \ui\tfa_{\lambda_1\lambda_2}(\bt;\om_1,\tht_1;\om_2\tht_2)
  \simeq \esp{\ui\ag 2\Delta(b)} \ampRid_{\lambda_1}(\bt;\om_1,\tht_1)
  \ampRid_{\lambda_2}(\bt;\om_2,\tht_2) \;.
\end{equation}
in terms of the one-graviton amplitude and of the elastic $S$-matrix.  It is
clear from eqs.~\eqref{ABCexpn} that such approximate relation neglects terms of
relative order $\ord{\hbar^2\om^2/E^2}=\ord{\om R/\ag}^2$, which are
negligible in the regime we are considering, and are subleading not only with
respect to terms $\sim\ag$ (like the eikonal phase $\delta(\bt)$) but also with
respect to the terms $\sim\om R$.

We expect an analogous factorization formula to hold for the generic
$N$-graviton emission amplitude off eikonal ladders (we explicitly checked the
3-graviton case), in the form
\begin{align}
  \ui\tfa(\bt;q_1,\cdots,q_N) \simeq  \esp{\ui\ag 2\Delta(b)}
  \prod_{r=1}^N \ampRid_{\lambda_r}(\bt;\om_r,\tht_r)\;.\label{ampNg}
\end{align}
Such an independent emission pattern corresponds to the final state
\begin{equation}\label{final}
  \ket{\text{gravitons; out}} = \sqrt{P_0}
  \exp\left\{\int\frac{\dif^3 q}{\hbar^3\sqrt{2\omega}}\;
    2\ui\sum_\lambda \ampRid_{\lambda}(\bt,\vq) a^\dagger_\lambda(\vq)
  \right\} \ket{0}
\end{equation}
in the Fock space of gravitons, with $P_0=1$, and creation
($a^\dagger_\lambda(\vq)$) and destruction ($a_\lambda(\vq)$) operators of
definite helicity $\lambda$ are normalized to a wave-number $\delta$-function
commutator $[a_\lambda(\vq),a^\dagger_{\lambda'}(\vq\,')]
=\hbar^3\delta^3(\vq-\vq\,')\delta_{\lambda\lambda'}$. However, this state
takes into account only real emission.  Virtual corrections can then be
incorporated by exponentiating both creation and destruction operators in a
(unitary) coherent state operator acting on the graviton vacuum $\ket{0}$ (the
initial state of gravitons). We thus obtain the full $S$-matrix
\begin{equation}\label{Shat}
  \hat{S}=\esp{2\ui\delta} \exp\left\{\int\frac{\dif^3 q}{\hbar^3\sqrt{2\omega}}\;
    2\ui\left[\sum_\lambda \ampRid_\lambda(\bt,\vq) a^\dagger_\lambda(\vq)
      +\text{h.c.}\right] \right\}
\end{equation}
that is unitary, because of the anti-hermitian exponent, when $b>b_c$.

By normal ordering eq.~\eqref{Shat} when acting on the initial state $\ket{0}$,
we find that the final state of graviton is still given by eq.~\eqref{final},
but with $P_0$ given by
\begin{equation}\label{P0}
  P_0 = \exp\left\{-2\int\frac{\dif^3 q}{\hbar^3\omega}\;
    \sum_\lambda |\ampRid_\lambda(\bt,\vq)|^2\right\} \;,
\end{equation}
which is just the no-emission probability, coming from the $a,a^\dagger$
commutators.

Due to the factorized structure of eq.~\eqref{final}, it is straightforward to
derive the inclusive distributions of gravitons and even their generating
functional
\begin{equation}\label{genFun}
  \mathcal{G}[z_\lambda(\vq)] = \exp\left\{2\int_{\Delta\omega}
    \frac{\dif^3 q}{\hbar^3\omega}\;\sum_\lambda|\ampRid_\bt^{(\lambda)}(\vq)|^2
    \left[ z_\lambda(\vq)-1\right] \right\} \;.
\end{equation}
In particular, the polarized energy emission distribution in the solid angle
$\Omega$ and its multiplicity density are given by
\begin{equation}\label{enDist}
  \frac{\dif E_\lambda^\GW}{\dif\omega\,\dif\Omega}
  = \hbar\omega\frac{\dif\Num_\lambda}{\dif\omega\,\dif\Omega}
  = 2\omega^2 \hbar |\ampRid_\lambda(\bt,\vq)|^2 \;, \quad
  \frac{\dif\Num}{\dif\om} = p(\om)
  = \frac1{\hom} \frac{\dif E^\GW}{\dif\omega}
\end{equation}
Both quantities will be discussed in the next section.

%===============================================================================
\subsection{Large $\bs{\om R}$ emission amplitude\label{s:lorea}}
%===============================================================================

In this section we analyse the graviton emission amplitude~\eqref{ampRid2} and
its spectrum~\eqref{enDist} generated by a small-angle ($\Theta_s\ll 1$)
scattering in the frequency region $\om\gtrsim R^{-1}$ and in the classical
limit $\hom \ll E$.

\begin{figure}[ht]
  \centering
  \includegraphics[width=0.49\linewidth]{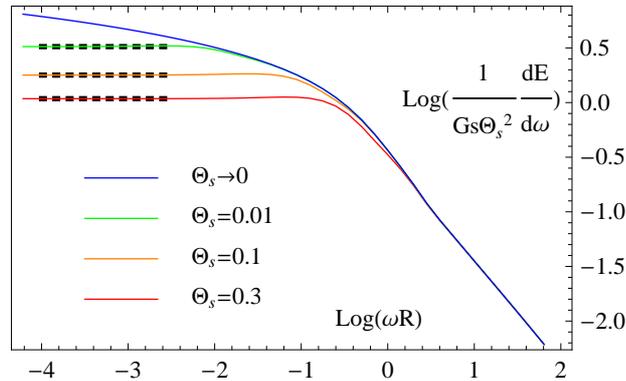}
  \caption{\it Frequency spectrum of gravitational
    radiation for various values of $\Theta_s$. For each $\Theta_s > 0$
    the ZFL value $\frac2{\pi} \log(1.65/\Theta_s)$ is obtained (dashed
    lines).}
\label{f:freqDist}
\end{figure}

We recall that the frequency spectrum integrated in the solid angle was already
studied in ref.~\cite{CCCV15} for large impact parameters $b\gg R$, i.e., for
small deflection angles $\Theta_s\ll 1$, both with and without rescattering
corrections. We briefly report the main results:
\begin{itemize}
\item For $\om R \lesssim \Theta_s$ the spectrum is flat and agrees with the
  zero-frequency-limit.
\item For $\Theta_s \lesssim \om R \lesssim 1$ the spectrum shows a slow
  (logarithmic) decrease with frequency. The behaviour in these two regions is
  rather insensitive to the inclusion of rescattering, and can be summarized by
  \begin{equation}\label{zfl}
  \frac{\dif E^\GW}{\dif\omega} \simeq Gs\Theta_E^2 \left[\frac{2}{\pi}
    \log\min\left(\frac1{\Theta_s},\frac1{\omega R}\right)
    + \text{const}\right] \qquad (\om R \lesssim 1)\;;
  \end{equation}
\item For $\om R \gtrsim 1$ the amplitude~\eqref{ampRid2} is dominated by
  small-$z$ values, and the $\zt$-integration can be safely extended to
  arbitrary large values without introducing spurious effects. The frequency
  distribution of radiated energy can then be well approximated by computing the
  square modulus of the amplitude~\eqref{ampRid2} by means of the Parseval
  identity, yielding
  \begin{equation}\label{spuv}
    \frac{\dif E^\GW}{\dif\omega}
    = 2 Gs \frac{\Theta_E^2}{\pi^2} \int\frac{\dif^2\zt}{|z|^4}\;\left(
      \frac{\sin\omega R \Phi(\zt)}{\omega R}\right)^2  \qquad (\om R\gtrsim 1)\;,
  \end{equation}
  whose asymptotic behaviour provides a spectrum decreasing like $1/\om$;
  more precisely
  \begin{equation}\label{invom}
    \frac{\dif E^\GW}{\dif\omega} \simeq Gs\Theta_E^2\frac1{\pi\om R}
    \qquad (\om R \gg 1) \;.
  \end{equation}
  In this region the inclusion of rescattering has the effect of lowering the
  spectrum by about 20\%. In any case, the total radiated-energy fraction up to
  the kinematical bound $\om_M=E/\hbar$ becomes
  \begin{equation}
    \label{kinBound}
    \frac{E^\GW}{\sqrt{s}} = \frac{\Theta_E^2}{2\pi}\log\ag
  \end{equation}
  and may exceed unity, thus signalling the need for energy-conservation
  corrections at sizeable angles (cf.\ sec.~\ref{s:far}).
\end{itemize}
In fig.~\ref{f:freqDist} we show the energy spectrum (divided by $Gs\Theta_E^2$)
for various values of $\Theta_s$. It is apparent that, for $\om R\gg 1$, its
shape is almost independent of $\Theta_s$. As we will show in sec.~\ref{s:rm},
there will be qualitative differences when approaching the strong-coupling region
$\Theta_s\sim 1$ where subleading contributions become important.

On the contrary, the angular distribution of graviton radiation studied in
ref.~\cite{CCCV15} didn't take into account rescattering contributions. The
latter are actually irrelevant for $\om R\ll 1$, but change drastically the
angular pattern for $\om R \gg 1$. In fact, the graviton exchanges between the
outgoing graviton $q$ and $p_2$ (see fig.~\ref{f:emissionRescatter2}) have the
main effect of deflecting the direction of $q$, just like the eikonal exchanges
between $p_1$ and $p_2$ are responsible of the deflection of $p_1$ (and $p_2$).
It turns out that the graviton radiation is collimated around the direction
$\Tht_s$ of the outgoing particle(s).

\begin{figure}[ht]
  \centering
  \includegraphics[width=0.25\textwidth]{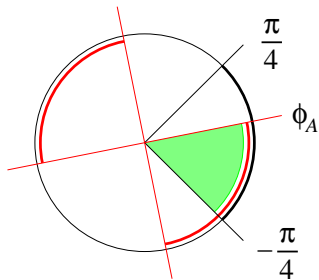}
  \caption{\it The end points in the integral~\eqref{zetaInt} correspond to the
    angular interval $[\phi_1,\phi_2]$ (green region), the latter being
    determined by the intersection of the $[-\pi/4,\pi/4]$ interval (black
    sector) with the region where $\sin(2(\phi_A-\phi))$ is positive (red
    sectors).}
  \label{f:phiSector}
\end{figure}

Quantitatively, the resummed emission amplitude~\eqref{ampRid2} in the classical
limit $\hbar\om\ll E$ and, say, for helicity $\lambda=-2$, reads
\begin{equation}\label{ResClas}
  \ampRid_\cl(\bt,\tht)
  = \sqrt{\ag} \frac{R}{\pi} \esp{-2\ui \phi_\tht} \int
  \frac{\dif^2 \zt}{2\pi {z^*}^2} \;
  \frac{\esp{\ui \omega b \zt \cdot (\tht- \Tht_s) }}{2\ui \omega R}
  \left( \esp{-2\ui \omega R \Phi_\cl(\zt)} -1 \right) \;,
\end{equation}
where $\Tht_s$ is the fast-particle scattering angle and
$\Phi_\cl$ was defined in eq.~\eqref{Phidef}.  We are interested in evaluating such
amplitude at large $\omega R$.  Since in the second exponential the function
\begin{equation}\label{quadPhi}
  \Phi_\cl(\zt) \equiv \hat{\bt}\cdot\zt + \log\left|\hat{\bt}-\zt\right|
 = \frac12 (z_2^2 -z_1^2) + \ord{|z|^3}
\end{equation}
vanishes (quadratically) at the origin, we expect that for
$\om R\gg 1$ the dominant contributions to the amplitude come from the small-$z$
integration region. By substituting the second-order expansion~\eqref{quadPhi}
into eq.~\eqref{ResClas} and by rescaling the integration variable
$\sqrt{\om R}\, z \equiv Z \equiv x+\ui y$, we obtain
\begin{equation}\label{defI}
  \frac{2\pi\om\ampRid}{\sqrt{\ag}\,\esp{-2\ui \phi_\tht}} \equiv I(\At) =
  \int\frac{\dif^2 Z}{2\pi Z^*{}^2} \; \esp{\ui 2\At\cdot\Zt}
  \left[\esp{\ui(x^2-y^2)}-1\right]\frac{1}{\ui} \;.
\end{equation}
which is a function of the 2-dimensional variable
\begin{equation}\label{aVar}
  \At \equiv |\At|(\cos\phi_A,\sin\phi_A)
  \equiv \sqrt{\om R}\, \frac{\tht-\Tht_s}{|\Tht_s|} \;,
  \qquad A\equiv |\At|\esp{\ui\phi_A} \in\C \;.
\end{equation}
Were it not for the factor $Z^*{}^2$ in the denominator, the r.h.s.\ of
eq.~\eqref{defI} would have the structure of a gaussian integral in 2
dimensions. It is possible however to provide a simple one-dimensional integral
representation for the function $I(\At)$ in eq.~\eqref{defI} (see
app.~\ref{a:angular}):
\begin{equation}\label{zetaInt}
  I(\At)=-\frac{A}{2A^*}\int_{\zeta_1}^{\zeta_2}\frac{\dif\zeta}{\zeta^2}\;
  \esp{-\frac{\ui}{2}A^*{}^2(\zeta^2+1)} \;,
\end{equation}
where the complex-integration endpoints $\zeta_l\equiv\esp{\ui 2\phi_l}:l=1,2$
are determined by the azimuth $\phi_A$, i.e., the azimuth of $\tht$ with respect
to $\Tht_s$, according to fig.~\ref{f:phiSector}. The function $I(\At)$
satisfies some
symmetry properties, and in particular it vanishes for $\phi=-\pi/4+n\pi: n\in\Z$.
This relation follows from the fact that, for
$\phi_A=(n-\frac14)\pi:n\in\Z$, the integration limits in eq.~\eqref{zetaInt}
coincide and thus the integral vanishes.

The intensity of the radiation on the tangent space of angular directions
centered at $\tht=\Tht_s$ and parametrized by $\At$ is shown in
fig.~\ref{f:azimQuadratic}.a. The main part of the radiation in the forward
hemisphere is concentrated around $|\At|\lesssim 1$, that means
$|\tht-\Tht_s|\lesssim |\Tht_s|/\sqrt{\om R}$, and is more and more collimated
around the direction $\Tht_s$ of the outgoing particle 1 for larger and larger
$\om R$.  This feature is a direct consequence of rescattering processes,
through which the emitted gravitons feels the gravitational attraction of
particle 2 and are therefore deflected, on average, in the same way as particle
1.

At given values of helicity and frequency, we observe a peculiar interference
pattern, with a vanishing amplitude at $\phi_A=\pm\pi/4+n\pi$ for helicity $\pm
2$. Such interference fringes are washed out when integrating the intensity over
some frequency range and summing over helicities. On the whole, the radiation
intensity is distributed almost isotropically around $\Tht_s$, with an
azimuthal periodicity (in $\phi_A$) resembling a quadrupolar shape.

This angular distribution differs from our prediction in~\cite{CCCV15}, where we
neglected rescattering and found graviton radiation distributed in the
scattering plane with angles ranging from $0$ (incoming particle 1) to
$\Tht_s$ (outgoing particle 1). By comparison, rescattering produces the above
dependence on $\tht-\Tht_s$, by associating in a clearer way jet 1 to the
outgoing particle 1.

Graviton radiation associated to large angle ($|\Tht_s|\sim 1$) scattering will
be analyzed in sec.~\ref{s:far} and compared to the previous one.

%%%%%%%%%%%%%%%%%%%%%%%%%%%%%%%%%%%%%%%%%%%%%%%%%%%%%%%%%%%%%%%%%%%%%%%%%%%%%%%%%
\section{Radiation model with ACV resummation\label{s:rm}}
%%%%%%%%%%%%%%%%%%%%%%%%%%%%%%%%%%%%%%%%%%%%%%%%%%%%%%%%%%%%%%%%%%%%%%%%%%%%%%%%%

In this section we extend the treatment of graviton radiation to scattering
processes characterized by large deflection angles $\Theta_s=\ord{1}$ or,
equivalently, to impact parameters $b\sim R$ of the order of the gravitational
radius, where the gravitational interaction becomes strong and a collapse is
expected to occur, at least at classical level. This requires to go beyond the
leading eikonal approximation reviewed in sec.~\ref{s:gbsa}, and to take into
account the nonlinear interactions which dominate at high energy. Such
corrections to the eikonal approximation have been identified~\cite{ACV90,ACV93}
and studied in detail for elastic scattering~\cite{ACV07,CC08,CC09,CC11}. Their
treatment is based on an effective action model that we are going to summarize
in sec.~\ref{s:ram} and to apply to graviton radiation in the rest of the
section.

%===============================================================================
\subsection{The reduced-action model\label{s:ram}}
%===============================================================================

The model consists in a shock-wave solution of the effective field theory
proposed by ACV~\cite{ACV93} in the regime $R\gg l_s$ of transplanckian
scattering on the basis of Lipatov's action~\cite{Li82}. The effective metric
fields of that solution have basically longitudinal ($h_{++}, h_{--}$) and
transverse ($h_{ij}:i,j=1,2$) components of the form
\begin{align}
  h_{--} &= 2\pi R a(\xt) \delta(x^-) \;, \qquad
  h_{++}=2\pi R \bar{a}(\xt)\delta(x^+) \;, \nonumber \\
  h &= \mathrm{Tr}(h_{ij}) = \bs{\nabla}^2\phi(\xt) \frac12 \Theta(x^+ x^-)
  \;,  \label{hij}
\end{align}
where we note wavefronts of Aichelburg-Sexl type~\cite{AiSe70} with
profile functions $a$ and $\bar{a}$ and an effective transverse field with
support in $x^+ x^- > 0$.

A simplified formulation of the solution~\eqref{hij} was obtained
in~\cite{ACV07} by an azimuthal averaging procedure which relates it to a
one-dimensional model in a transverse radial space with the axisymmetric action
\begin{equation}\label{ar}
  \ac = 2\pi^2 G s \int\dif r^2\left[\bar{s}a+s\bar{a} - 2\rho \dot{a}
  \dot{\bar{a}} - \frac{2}{(2\pi R)^2}(1-\dot{\rho})^2\right]
  \qquad \left(\cdot \equiv \frac{\dif}{\dif r^2} \right)
\end{equation}
in which $r^2$ plays the role of time parameter. Here $\phi(r^2)$ is replaced by
the auxiliary field $\rho(r^2)$ --- a sort of renormalized squared-distance ---
defined by
\begin{equation}\label{rho}
  \rho \equiv r^2 [1-(2\pi R)^2\dot\phi] \;, \qquad
  h \equiv \bs{\nabla}^2\phi = 4\frac{\dif}{\dif r^2}(r^2\dot\phi)
  = \frac{1}{(\pi R)^2}(1-\dot\rho) \;,
\end{equation}
which incorporates the basic $\phi,a,\bar{a}$ interaction, with effective
coupling $R^2$. Furthermore, the axisymmetric sources $s(r^2)=\delta(r^2)/\pi$
and $\bar{s}(r^2)=\delta(r^2-b^2)/\pi$ describe (approximately) the energetic
incident particles and $\phi(r^2)$ is taken to be real-valued --- as for the TT
polarization only --- thus neglecting the infrared singular one in the frequency
range $\om\sim 1/R$ we are interested in.

The equations of motion of~\eqref{ar} for the profile functions admit two
constants of motion, yielding the relations
\begin{equation}\label{eoma}
  \dot{a} = -\frac1{2\pi\rho} \;, \qquad
  \dot{\bar{a}} = -\frac1{2\pi\rho}\Theta(r^2-b^2) \;,
\end{equation}
while that for the field $\rho$ yields
\begin{equation}\label{eomrho}
  \ddot{\rho} = 2(\pi R)^2 \dot{a}\dot{\bar{a}} =
  \frac{R^2}{2\rho^2}\Theta(r^2-b^2) \;, \qquad
  \dot{\rho}^2+\frac{R^2}{\rho} = 1 \qquad (r > b) \;.
\end{equation}
The latter describe the $r^2$-motion of $\rho(r^2)$ in a Coulomb field, which is
repulsive for $\rho>0$, and acts for $r^2>b^2$ only, so that $b^2$ actually cuts
off that repulsion in the short-distance region.

The interesting solutions of~\eqref{eoma} and~\eqref{eomrho} are those which are
ultraviolet safe --- for which the effective field theory makes sense --- and
are restricted by the regularity condition $\rho(0)=0$ which avoids a possible
$r^2=0$ singularity of the $\phi$-field.

External ($r>b$) and internal ($0<r<b$) regular solutions are easily written
down for this solvable model
\begin{align}
  \rho &=
  \begin{cases}
    R^2\cosh^2\chi(r^2) & (r^2 \geq b^2) \\
    \rho(b^2)+\dot\rho(b^2)(r^2-b^2) & (0 \leq r^2 \leq b^2)
  \end{cases} \label{solrho} \\
  r^2 &= b^2 + R^2 (\chi + \sinh\chi\cosh\chi -\chi_b -\sinh\chi_b\cosh\chi_b)
  \qquad \left(\chi_b\equiv\chi(b^2)\right)
  \nonumber
\end{align}
and are matched at $r^2=b^2$ by the condition ($t_b\equiv\tanh\chi_b$)
\begin{equation}\label{mc}
  \rho(b^2) = R^2\cosh^2\chi_b = b^2\dot\rho(b^2) = b^2 t_b\;,
  \qquad \frac{R^2}{b^2} = t_b(1-t_b^2) \;.
\end{equation}

The criticality equation~\eqref{mc} is cubic in the $t_b$ parameter and
determines the branches of possible solutions with $\rho(0)=0$. For
$b^2>b_c^2\equiv(3\sqrt{3}/2)R^2$ there are two real-valued, non-negative
solutions, and the ``perturbative'' one with $t_b\to 1$ for $b\gg b_c$ is to be
taken. By replacing such solution in the action~\eqref{ar} we get the
non-perturbative on-shell expression
\begin{align}
  2\delta(b,s) \equiv \ac &= \ag\int_0^{L^2}\frac{\dif r^2}{R^2}\;,\left[
    \frac{R^2}{\rho}\Theta(r^2-b^2)-(1-\dot\rho)^2\right] \nonumber \\
  &= \ag \left[2\chi_L-2\chi_b+1-\frac1{t_b}\right] \qquad (b > b_c)
  \;, \label{osact}
\end{align}
where $L$ is an IR cutoff needed to regularize the Coulomb singularity.
The phase-shift~\eqref{osact} shows the large-$b$ behaviour
\begin{equation}\label{deltaLargeb}
  \delta(b,s)\simeq \ag\left(\log\frac{L}{b}+\frac{R^2}{4b^2}+\cdots\right) \;,
\end{equation}
which however is only qualitatively correct for the subleading term whose full
expression is actually~\cite{ACV90}
\begin{equation}\label{redh}
  \Re\delta_H = \delta(b,s)-\ag\log\frac{L}{b} = \ag\frac{R^2}{2b^2} \;.
\end{equation}
The difference is due to the various approximations being made (one polarization
and azimuthal averaging).

\begin{figure}[ht]
  \centering
  \includegraphics[width=0.7\linewidth]{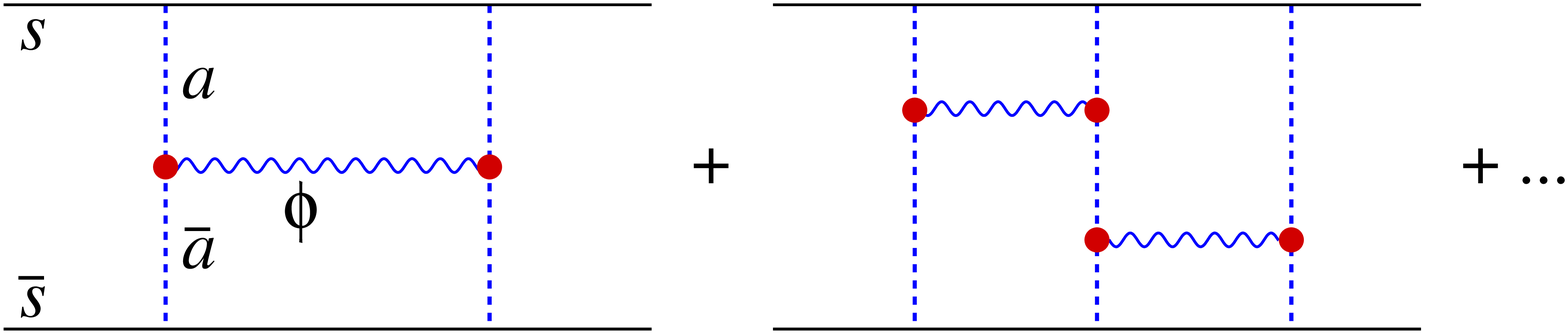}
  \caption{\it The H diagram (left) and the first multi-H diagram (right)
    starting the series of subleading contributions to the eikonal.}
  \label{f:multiH2}
\end{figure}

Despite such approximations, the importance of the non-perturbative
expressions~\eqref{solrho} and \eqref{osact} for solutions and action is to
provide a resummation of all subleading contributions $\sim(R^2/b^2)^n$ to the
eikonal of multi-H type (fig.~\ref{f:multiH2}) and to exhibit its singularity
structure in the classical collapse regime, on the basis of the criticality
equation~\eqref{mc}.

In fact, for $b<b_c$ we find that no real-valued solutions exist and $t_b$
acquires an imaginary part. The solution with negative $\Im t_b$ has
$\Im\ac >0$, is stable and is close to the perturbative solution at large
distances. The corresponding action is found to yield a suppression of the
elastic channel of type
\begin{equation}\label{elsup}
  |S_\el(b,s)|^2 \simeq
  \exp\left\{-\frac{4\sqrt{2}}{3}\ag\left(1-\frac{b^2}{b_c^2}
      \right)^{3/2}\right\} \;,
\end{equation}
which can be related to a tunnel effect~\cite{CC08,CC09}
through the repulsive Coulomb-potential barrier which is classically forbidden.

Actually, the action shows a branch-point singularity at $b=b_c$ of index $3/2$
with the expansion
\begin{align}
  \ac-\ac_c &= \ag \left[\sqrt{3}\left(1-\frac{b^2}{b_c^2}\right)
    \pm\ui\frac{2\sqrt{2}}{3}\left(1-\frac{b^2}{b_c^2}\right)^{3/2}+\cdots \right]
    \;, \nonumber \\
    t_b &= \frac1{\sqrt{3}} \mp\ui\frac{\sqrt{2}}{3}
    \sqrt{1-\frac{b^2}{b_c^2}} + \cdots \label{acrit}
\end{align}
which is thus responsible for the suppression~\eqref{elsup} just mentioned. The
presence of the index $3/2$ seems a robust feature of this kind of models
because the expansion of the action in $t_b$ starts at order
$(t_b-1/\sqrt{3})^2$, due to the action stationarity, thus avoiding a
square-root behaviour.

The result so obtained is puzzling, however, because it may lead to unitarity
loss~\cite{CC09,CC11}, unless some additional state, or radiation enhancement, is found in the
$b\leq b_c$ region. In fact, it represents a basic motivation of the present
paper, and of the following treatment of the radiation associated to the ACV
resummation.

%===============================================================================
\subsection{Single graviton emission by H-diagram exchange}
%===============================================================================

\begin{figure}[ht]
  \centering
  \includegraphics[width=0.7\linewidth]{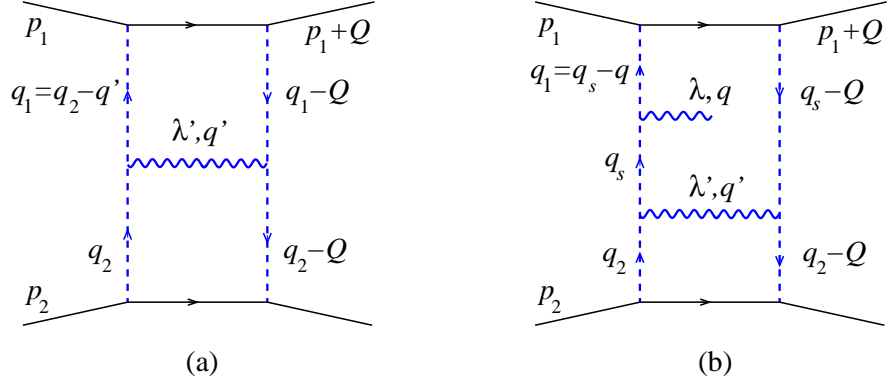}
  \caption{\it Emission from H diagram.}
  \label{f:HdiagEmission}
\end{figure}

Here we want to argue that the graviton radiation associated to the H-diagram
eikonal exchange is well described by a generalization of the soft-based
representation in eq.~\eqref{Mmatch}. To this purpose we shall use the
dispersive method of~\cite{ACV90}, which consists in relating both (a) exchange
and (b) emission to the multi-Regge amplitudes~\cite{Li82,CCV15} pictured in the
overlap functions of fig.~\ref{f:HdiagEmission}.

For the H-diagram (fig.~\ref{f:HdiagEmission}.a), the CCV helicity
amplitude~\cite{CCV15} for emitting a graviton of momentum $\qt'=\qt_2-\qt_1$
and helicity $\lambda'$, in the center of mass frame with incident momentum
along the $\hat z$-axis, is given by [cf.\ eq.~\eqref{Mregge}]
\begin{align}
  \amp_\regge(\qt',E,\om,\qt_2) &= \frac{\kappa^2 s^2 J_L^{\mu\nu}
    \pol_{\mu\nu}^{(\lambda')}}{(\qt-\qt_2)^2 \qt_2^2}
  = \frac{\kappa^3 s^2}{\qt'{}^2}F^{(\lambda')}(\qt_2,\qt_2-\qt')
  \nonumber \\
  F^{(\lambda)}(\qt_2,\qt_1) &\equiv
  1-\esp{\ui\lambda( \phi_{\qt_2}-\phi_{\qt_1})} \;. \label{Hregge}
\end{align}
Correspondingly, the overlap-function (fig.~\ref{f:HdiagEmission}.a), at
generally non vanishing momentum transfer $\Qt$, and for incidence direction
along the $\hat z$-axis in the center of mass frame, is proportional to the
Lipatov graviton kernel~\cite{Li82}
\begin{equation}\label{kernel}
  K(\qt_2,\qt_1;\Qt) \equiv \sum_{\lambda'}
  J_L^{\mu\nu}(\qt_1,\qt_2)\pol_{\mu\nu}^{(\lambda')}(\qt')
  J_L^{\mu'\nu'}(\qt'_1,\qt'_2)\pol_{\mu'\nu'}^{(\lambda')*}(\qt') \;,
\end{equation}
where $J_L$ is the Lipatov current~\cite{Li82} and
$\qt'_i\equiv\qt_i-\Qt:i=1,2$. In two transverse dimensions, where the $\qt_i$'s
are all coplanar, the explicit result is
\begin{equation}\label{K2d}
  K(\qt_2,\qt_1;\Qt) =
  \frac{4\,\qt_1^2\,\qt_2^2\,\qt'_1{}^2\,\qt'_2{}^2}{[(\qt_1-\qt_2)^2]^2}
  2\sin\phi_{12}\sin\phi_{1'2'}\cos(\phi_{12}-\phi_{1'2'})
\end{equation}
and checks with ref.~\cite{Li82}.

The result~\eqref{K2d} is valid for on-shell intermediate particles, and provides
directly, by integration over $\qt_i$ and Fourier transform in $\Qt$ to
$\bt$-space, the imaginary part of the H-diagram amplitude, or~\cite{ACV07}
\begin{equation}\label{ImDeltaH}
  \Im\delta_H(b,s) \equiv Y Gs R^2 \int \dif^2\qt'\;
  \left|\tilde{h}(\bt,\qt')\right|^2 \;,
\end{equation}
where
\begin{equation}\label{htilde}
  \tilde{h}(\bt,\qt) = 2 \int\frac{\dif^2\qt_2}{(2\pi)^2} \;
  \frac{\esp{\ui\bt\cdot\qt_2}}{|\qt|^2}
  \left[1-\esp{2\ui(\phi_{\qt_2-\qt}-\phi_{\qt_2})}\right] \;,
\end{equation}
is the $h$-field in $\qt$-space at $\lambda'=-2$~\cite{CCV15,CCCV15}.%
\footnote{The quantities $h = \mathrm{Tr}(h_{ij}) \rightsquigarrow h_{zz^*}$ and
  $h_s\rightsquigarrow h_{zz},h_{z^* z^*}$ are related to different components
  of the metric fields $h_{\mu\nu}\equiv g_{\mu\nu}-\eta_{\mu\nu}$ in the
  shock-wave solution~\eqref{hij}. For a more precise identification,
  see~\cite{CCCV15}.}
The quantity~\eqref{htilde} has a logarithmic divergence, because of the known
residual infrared singularity $\sim 1/\qt'{}^2$ of the integrand
in~\eqref{ImDeltaH}, due to the LT polarization. Such divergence is expected and
is compensated in observables by real emission in the usual way, so as to lead
to finite, but resolution dependent results.

On the other hand, we are looking for $\Re\delta_2$, the H-diagram contribution
to the 2-loop eikonal, which is supposed to be IR safe, because a
$\bt$-dependent IR divergence would be observable, and inconsistent with the
Block-Nordsieck factorization theorem. In~\cite{ACV90} it was shown that fixed
order dispersion relations plus $S$-matrix exponentiation lead indeed to the
finite result
\begin{equation}\label{ReDeltaH}
  \Re\delta_2(b,s) =  \left.\frac{\pi}{2Y}\Im\delta_H(b,s)\right|_{\mathrm{Reg}}
  = \frac{\pi}{2} Gs R^2
  \int \dif^2\qt'\;\left|\tilde{h}(\bt,\qt')\right|^2_{\mathrm{Reg}}
  \stackrel{b\gg R}{=} \ag\frac{R^2}{2b^2} = \frac{2G^3 s^2}{b^2} \;.
\end{equation}
Here the regularization subtraction is due to the second order contributions of
$\delta_0$ and $\delta_1$ to the $S$-matrix exponential.

Our present purpose is actually to compute the graviton radiation associated to
the H-diagram, in which a further Regge graviton vertex is introduced in all
possible ways, as in fig~\ref{f:HdiagEmission}.b for the upper-left corner. In
the limit $|\qt|=\hom \ll E$ --- that we assume throughout the paper --- the
dominant contributions are for $|\qt| \ll |\qt'| \sim m_P$, so that no
insertions on the $\qt'$-exchange should be considered. As a consequence, for
$q$ in jet 1 and using the CCV gauge~\cite{CCV15} in which jet 2 is switched
off, only the upper-left and upper-right insertions will be considered.

Consider first the upper-left diagram (fig.~\ref{f:HdiagEmission}.b) at
imaginary part level.  For any fixed values of $\qt$, $\qt'$ and $\Qt$, the
integrand has the form
\begin{align}
  \sqrt{\ag} R \int\dif^2\qt_s\;\esp{\ui\Qt\cdot\bt}
  & \left(\frac1{\qt'{}^2}\right)^2 F^{(\lambda')}(\qt'_s+\qt',\qt'_s)
  F^{(\lambda')}(\qt_s+\qt',\qt_s) \nonumber \\
  &\times\kappa\frac{|q_s|^2}{|q|^2}\left(1-\frac{q_s^*}{q_s}
    \frac{q_s-q}{q_s^*-q^*} \right)  + \cdots \;,\label{Hda}
\end{align}
where $\qt_s\equiv \qt_2-\qt'$, $\qt'_s=\qt_s-\Qt$, and we have taken, for
definiteness, $\lambda = -2$. The $|q_s|^2$ factor in the numerator is needed in
order to have the proper counting of $|q_i|^2$ denominators in multi-Regge
factorization~\cite{Li82}.

\begin{figure}[ht]
  \centering
  \includegraphics[width=0.88\linewidth]{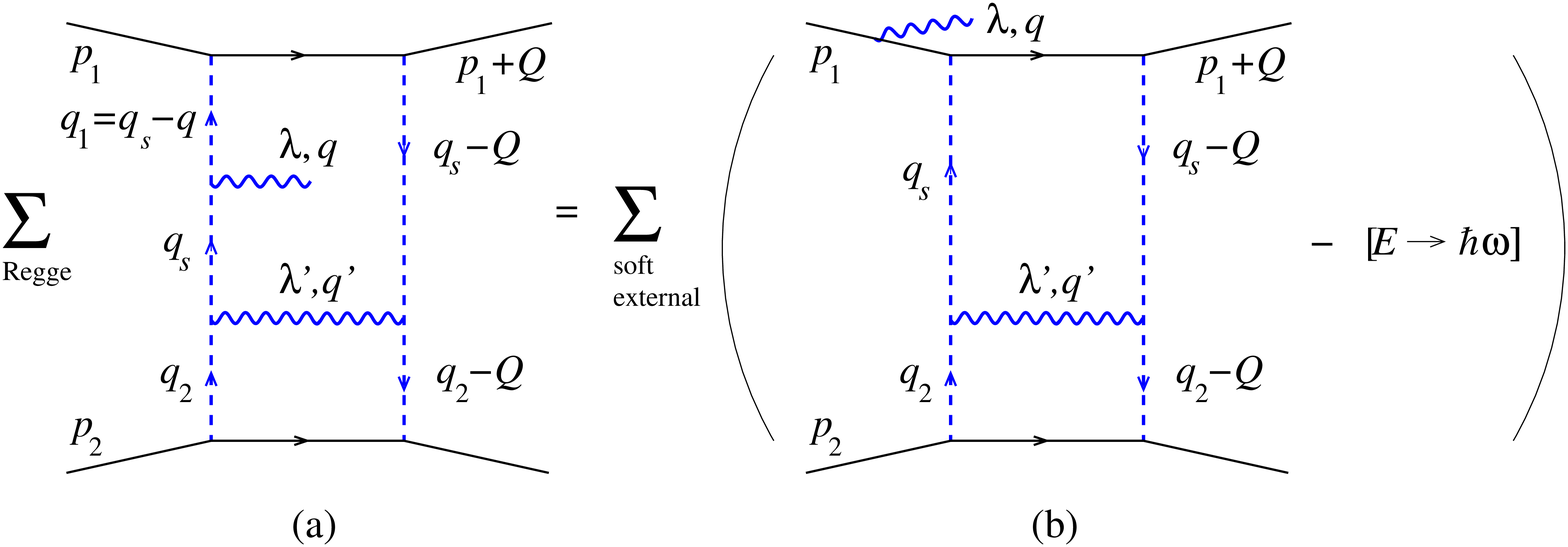}
  \caption{\it Diagrammatic representation of the soft-based emission amplitude.}
  \label{f:softBasedHdiag}
\end{figure}

We then apply to eq.~\eqref{Hda} the same reasoning used in sec.~\ref{s:uea} to
match the soft and Regge limits. By the approximate identity
\begin{equation}\label{appId}
  \frac{|q_s|^2}{|q|^2} \left(1-\frac{q_s^*}{q_s}\frac{q_s-q}{q_s^*-q^*}\right)
  \simeq \frac{q q_s^* - q^* q_s}{q}
  \left[\frac1{q^*-\omE q_s^*}-\frac1{q^*-q_s^*} \right] \;,
\end{equation}
valid in the region $(\hom/E)|q_s|\ll |q|$, we derive the relationship between Regge
and soft insertion analogous to eqs.~\eqref{Mmatch} and~\eqref{softRegge}
\begin{equation}\label{reggeSoft}
  \text{Regge}|_E = \text{soft}|_E - \text{soft}|_{\hom} \;,
\end{equation}
in which

\begin{equation}\label{softE}
  \text{soft}|_E = \frac{\hom}{E}
  [\esp{2\ui(\phi_{\qt-\frac{\hom}{E} \qt_s}-\phi_{\qt})}-1]
\end{equation}
for the upper-left case. A similar relationship holds for the upper-right case
and, by including both jets, for any soft insertions, as pictured in
fig.~\ref{f:softBasedHdiag}.

We note at this point that in eq.~\eqref{reggeSoft} the insertion factor for the
intermediate particle $p_1+q_s$ cancels out between left and right insertions,
so that we get in total the insertion factor for external legs only, in the form
\begin{align}
  &\kappa\left[\frac{E}{\hom}\left(\frac{q^*}{q}\frac{q-\omE Q}{q^*-\omE Q^*}-1
    \right)-\{E\to\hom\}\right] \nonumber \\
  &\qquad\qquad =\kappa\left[ \frac{E}{\hom}\left(
    \esp{2\ui(\phi_{\qt-\frac{\hom}{E} \Qt}-\phi_\qt)}-1\right)
  -\left(\esp{2\ui(\phi_{\qt-\Qt}-\phi_\qt)}-1\right)\right] \;. \label{extLeg}
\end{align}
The latter replaces in eq.~\eqref{Hda} the sum of Regge insertions and is
only dependent on the overall momentum transfer $Q$
(fig.~\ref{f:softBasedHdiag}).

Since the above factorization in $\Qt$-space holds for any fixed values of $\Qt$
and $|\qt'|>|\qt|$, it is presumably valid for the IR regularization procedure
of eq.~\eqref{ReDeltaH} also, because the latter consists in subtracting the IR
singularity due to lower order eikonal contributions to the $S$-matrix
exponential. We shall then assume eq.~\eqref{extLeg} for the full graviton
emission amplitude associated to H-diagram exchange. This leads to the
expression
\begin{align}
  \tfa_H(\bt,E,\qt) &= \sqrt{\ag}\frac{R}{\pi}\int\frac{\dif^2\Qt}{2\pi}\;
  \tilde{\Delta}_H(\Qt)\esp{\ui\Qt\cdot\bt} \nonumber \\
  &\quad\times \left[\frac{E}{\hom}\left(\esp{2\ui(\phi_{\qt-\frac{\hom}{E}\Qt}
        -\phi_\qt)}-1\right)-\left(\esp{2\ui(\phi_{\qt-\Qt}-\phi_\qt)}-1
      \right)\right] \;, \label{tfaH}
\end{align}
where $\ag\tilde{\Delta}_H(\Qt)$ is the (regularized) inverse Fourier transform
of $\Re\delta_2(b)$ in eq.~\eqref{ReDeltaH}.

The main achievement of eq.~\eqref{tfaH} is its independence of the detailed
structure of the H-diagram because of the factorization of the soft insertions
in $\Qt$-space. Therefore, it is the generalization of the soft-based
representation of the unified amplitude to the next-to-leading eikonal exchange.

%===============================================================================
\subsection{Soft-based representation and eikonal resummation\label{s:sbr}}
%===============================================================================

We have just argued that the single graviton emission amplitude associated to
H-diagram exchange is provided by eq.~\eqref{tfaH} which is directly expressible
in terms of the H-diagram amplitude in $\Qt$-space. We can even use the
$z$-representation for the phase transfers
\begin{equation}\label{zRep}
  \esp{2\ui \phi_{\tht}} - \esp{2\ui \phi_{\tht'}}
  = -2 \int \frac{\dif^2\zt}{{2\pi z^*}^2} \left( \esp{\ui A \zt \cdot \tht}
    - \esp{\ui A \zt \cdot \tht'} \right) \;, \qquad(A\in \R^*)
\end{equation}
and, by exchanging the order of $\Qt$- and $\zt$-integrals we recast
eq.~\eqref{tfaH} in the form
\begin{equation}\label{Mres}
  \tfa = \sqrt{\ag}\frac{R}{\pi}\esp{-2\ui\phi_\tht}\int
  \frac{\dif^2\zt}{2\pi z^{*2}}\;\esp{\ui b\zt\cdot\qt} \left\{\frac{E}{\hom}
    \left[\Delta(\bt-\omE b\zt)-\Delta(\bt)\right]-\left[\Delta(\bt-b\zt)
      -\Delta(\bt)\right]\right\} \;,
\end{equation}
where $\Delta=\Delta_0+\Delta_H$ and $\Delta_H(\bt)=R^2/2b^2$, thus generalizing
the soft-based representation of eq.~\eqref{Mmatch} to the next-to-leading (NL)
term. We shall base on eq.~\eqref{Mres} the subsequent formulation of our
radiation model.

We note immediately, however, that eq.~\eqref{Mres} has a purely formal meaning
in the region $|b\zt-\bt|=\ord{R}$, because the H-diagram
expression~\eqref{ReDeltaH} breaks down whenever $R/b$ is not small. We are thus
led to think that we have to know something about the behaviour of $\Delta(\bt)$
in the large-angle regime $b\sim R$ before even writing the
representation~\eqref{Mres} we argued for.

That is precisely what the reduced action model --- as summarized in
sec.~\ref{s:ram} --- provides for us. Indeed it consists in the resummation of
the multi-H diagrams (fig.~\ref{f:multiH2}) of the eikonal, which is the set of
two-body irreducible diagrams without a rescattering subgraph. Such diagrams are
expected to share with the NL term the property that the central subgraphs have
energetic $q'$-type exchanges, where $|q'|$ is of the order of the Planck mass
or larger, thus suppressing their contribution to soft insertions with
$\om\sim R^{-1}\sim m_P^2/E$.

For that reason we think, we can repeat the argument with peripheral Regge
insertions elaborated before, and then use the soft-Regge
identities~\eqref{reggeSoft} to derive the external-particles insertion
formula~\eqref{tfaH} and the soft-based representation~\eqref{Mres}. As a
result, we are now able to look at the integrals in eq.~\eqref{Mres} in a
realistic way by setting
\begin{equation}\label{DeltaRes}
  2\Delta(\bt) = \frac{2\delta(\bt)}{\ag} = -2\chi_b+1-\frac1{t_b} \;,
\end{equation}
where $\delta(\bt)$ is the irreducible eikonal function with ACV resummation
(after factorization of the IR part $\sim\log L/b$), which extrapolates the NL behaviour to
small values of $b\sim R$. The latter is given in terms of the
solution~\eqref{osact} for the reduced-action model action, and
$t_b=\tanh\chi_b$ and $\chi_b$ are determined by the matching condition of
eq.~\eqref{mc}.

The expressions~\eqref{DeltaRes} and~\eqref{osact} are now well-defined for
$b^2 > b_c^2 = \frac{3\sqrt{3}}{2}R^2$, where the role of the singularity at
$b=b_c$ will be discussed soon. The result~\eqref{Mres} contains the resummed
modulating function
\begin{align}
  \Phi_\ram(\om,\zt) &\equiv \frac{E}{\hom}\left[\Delta(\bt-\omE b\zt)-\Delta(\bt)
    \right] -\left[\Delta(\bt-b\zt)-\Delta(\bt)\right] \nonumber \\
    &\stackrel{\hom\ll E}{\simeq} -b \Delta'(b) \hat{\bt}\cdot\zt +\Delta(\bt)-\Delta(\bt-b\zt)
   \equiv\Phi_{\ram,\cl}(\zt) \label{PhiRAM}
\end{align}
(yielding $\Phi_\cl$ of eq.~\eqref{Phidef} in the classical limit and large-$b$
region), which generalizes the expressions~\eqref{PhiRdef} and~\eqref{denom1}
for the leading term, and enters the corresponding soft field
\begin{equation}\label{hsRAM}
  h_s^{(\lambda)}(\om,z) \equiv
  -\frac{\Phi_\ram(\om,\zt)}{\pi^2\zl} \;.
\end{equation}

Next step is to sum up all single-graviton emission amplitudes from any of the
$\bk{n}\sim\ag\gg1$ irreducible eikonal exchanges with ACV resummation, by
taking into account two important effects: (a) The correct phase and
$\qt$-dependence for all various incidence angles and (b) the rescattering of
the emitted graviton with the fast particles themselves.

Both effects can be taken into account by the generalized $\bt$-space
factorization formula explained in sec.~\ref{s:eer} for the leading graviton
exchange. By replacing $\tfa_\el = 2\delta_0$ by $2\delta$, the resummed soft
field of eq.~\eqref{sum1} becomes
\begin{align}
  &\frac1{z^{*2}}\Phi_\ram(\om,\zt) \frac{\esp{2\ui\delta(\bt-\omE b\zt)}-
    \esp{2\ui\left[\delta(\bt)+\omE\left(\delta(\bt-b\zt)-\delta(\bt)\right)
      \right]}}{2\ui\left[\delta(\bt-\omE b\zt)-\delta(\bt)-\omE
    \left(\delta(\bt-b\zt)-\delta(\bt)\right)\right]} \nonumber \\
 &=\frac1{z^{*2}}\frac{\esp{2\ui\delta(\bt)}}{2\ui\om R}\left[
   \esp{2\ui\om R \frac{E}{\hom}\left[\Delta(\bt-\omE b\zt)-\Delta(\bt)\right]}
   -\esp{2\ui\om R\left[\Delta(\bt-b\zt)-\Delta(\bt)\right]}\right] \;,
\end{align}
where we have canceled out the $\Phi_\ram$ function at numerator with the same
factor in the denominator, and factored out the eikonal $S$-matrix
$\esp{2\ui\delta(\bt)}$.

By applying the definition~\eqref{ampRid2}, the full graviton emission
probability amplitude becomes
\begin{align}
  \frac{\ampRid_\lambda(\bt;\om,\tht)}{\esp{\ui\lambda\phi_\tht}} &=
  \sqrt{\ag}\frac{R}{\pi}\int\frac{\dif^2\zt}{2\pi|z|^2\esp{\ui\lambda\phi_z}}
  \frac{\esp{\ui b\zt\cdot\qt}}{2\ui\om R}\left\{\esp{2\ui\om R\left[
        \Delta(\bt-b\zt)-\Delta(\bt)\right]}-\esp{2\ui\om R\frac{E}{\hom}\left[
        \Delta(\bt-\omE b\zt)-\Delta(\bt)\right]} \right\} \nonumber \\
  &\simeq \sqrt{\ag}\frac{R}{\pi}\int\frac{\dif^2\zt}{2\pi|z|^2\esp{\ui\lambda\phi_z}}
  \frac{\esp{\ui b\om\zt\cdot(\tht-\Tht_s)}}{2\ui\om R} \left(
    \esp{-2\ui\om R\Phi_{\ram,\cl}(\zt)}-1\right) \;, \label{ampRis}
\end{align}
where $\Tht_s(\bt)\equiv -b\Delta'(b)\Tht_E=\Tht_E/t_b$,
$\Tht_E\equiv-(2R/b)\hat{\bt}$ and $\Phi_{\ram,\cl}$ is the classical limit of
$\Phi_\ram$ introduced in eq.~\eqref{PhiRAM}.

We note that the $\om R$-dependent correction factor to naive
$\bt$-factorization takes into account in a simple and elegant way both
incidence angle dependence and elastic rescattering with the incident particles
including ACV resummation too.

\begin{figure}[ht]
  \centering
  \includegraphics[width=0.22\linewidth]{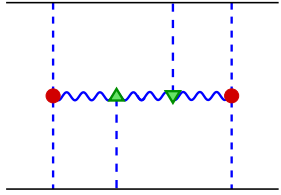}
  \caption{\it The first rescattering diagram contributing to the eikonal phase.}
  \label{f:rescatteringDiagram}
\end{figure}

One may wonder at this point about the role of the rescattering contributions to
the irreducible eikonal not included here in $\delta(\bt)$, and starting at
order $R^4/b^4$ (fig.~\ref{f:rescatteringDiagram}). The latter presumably have a
massless 3-body discontinuity and have thus the interpretation of $2\to 3\to 2$
transition in the rescattering process, leading to a recombination in a 2-body
state. This would imply taking into account inelastic higher-order contributions
to rescattering, a feature which is outside the scope of the present paper.

%===============================================================================
\subsection{Coherent state and correlation effects\label{s:ce}}
%===============================================================================

The derivation of the coherent-state operator proceeds now as in
sec.~\ref{s:mge} if we stick to the ``linear'' approximation, which neglects
correlation effects. The only difference is the replacement of $\delta_0(b)$ by
$\ag\Delta(\bt)$ in the amplitude $\ampRid_\lambda(\bt,\vq)$ of
eq.~\eqref{ampRis}, so that we obtain
\begin{equation}
  \hat{S} = \esp{2\ui\delta(\bt)} \exp\left\{\int\frac{\dif^3 q}{\sqrt{2\om}}\;
    2\ui\left(\sum_\lambda\ampRid_\lambda(\bt,\vq) a_\lambda^\dagger(\vq)
      +\text{h.c.}\right)\right\} \;. \label{rcs}
\end{equation}

We shall base on eq.~\eqref{rcs} most of the subsequent results. But we
want to provide a preliminary discussion of the limits of that approximation and
of the size of correlations that we can envisage. That is important for the
ACV-resummed model, because we would like to describe sizeable scattering angles
$\Theta_s\sim R/b \sim\ord{1}$, while approaching the collapse regime.

We start noticing that many-body correlations are already present from start in
the many-graviton states of sec.~\ref{s:mge} and are in principle calculable. For
instance, the 2-body correlation can be estimated from eq.~\eqref{ABCexpn} and is,
order of magnitude like,
\begin{equation}\label{c12}
  c_{12} \sim \ampRid_1 \ampRid_2 \big( |z_1|\om_1 R |z_2| \om_2 R\big)/\ag \;.
\end{equation}
We note also that the factors $|z_i|=|x_i/b|\sim1/\sqrt{\om_i R}$ are small in
the dominant integration region for radiation (sec.~\ref{s:lorea}) so that
$c_{12}$ becomes of relative order $\sqrt{\om_1\om_2}/E\simeq 1/\ag\ll 1$ for
$\om_i R\sim\ord{1}$. This means that, within our assumptions, we can neglect
finite order correlations.

One may wonder, however, whether correlated emission can be enhanced by
multiplicity effects --- not only those of the exchanged gravitons
($\bk{n}\sim\ag$) --- but also those of the emitted ones
($\bk{N}\sim\ag\Theta_s^2$), a number which may be large, and even more for
$\Theta_s=\ord{1}$.

One such effect is certainly present, and is due to energy conservation. Even if
energy transfer is explicitly considered in the treatment of rescattering in
secs.~\ref{s:eer} and \ref{s:mge}, the kinematical constraints are not
explicitly enforced. But such constraints are needed, because the expected
average emitted energy $\bk{\om}\equiv E/\bk{N}=R^{-1}\Theta_s^{-2}$ is of the
order of the so-called classical cutoff~\cite{GrVe14,CCCV15} and cannot be large
if $\Theta_s$ increases up to $\ord{1}$. This means that the larger values of
$\om R$ can be reached only for a smaller number of gravitons, thus distorting
the calculation of inclusive distributions. That effect is therefore important,
but can be included in the coherent state~\eqref{rcs} and will be discussed in
sec.~\ref{s:ecc}.

Another kind of multiplicity effect --- not included in~\eqref{rcs} ---
comes from multi-graviton emission by a single exchange. A simple model for that
is to consider soft emission which, according to~\cite{ACV90}, sec.~4, is
described by the operator eikonal
\begin{align}
  \hat\delta_{\text{soft}}(\bt,a_\qt) &= \ag\int\frac{\dif^2\qt_s}{(2\pi)^2}\;
  \frac{\esp{\ui\qt\cdot\bt}}{\qt_s^2}\Usoft^{\qt_s}(a_\qt) \nonumber \\
  \Usoft^{\qt_s}(a_\qt) &\equiv \exp\left\{
    2\sqrt{G}\int\frac{\dif^3 q}{\sqrt{2\om_q}}
    |\qt_s|\frac{\sin(\phi_\qt-\phi_{\qt_s})}{|\qt|b}
    \left[a_\lambda^\dagger(\qt)-a_\lambda(\qt)\right]
  \right\} \;. \label{Usoft}
\end{align}
Here we can see the nonlinear structure of the operator~\eqref{Usoft} as
``coherent state of coherent states'' in the soft limit. Its linear part agrees
with the state~\eqref{rcs} by the approximate form of
\begin{equation}\label{ampridb}
  \ampRid \simeq \sqrt{\ag} \frac{\Theta_E}{2\pi}
  \frac{\sin(\phi_\qt-\phi_{\qt_s})}{|\qt|} J_0(|\qt|b) \;,
\end{equation}
which is valid in the region $(E/\hom)|\qt|\gg |\qt_s|\gg|\qt|$~\cite{CCCV15}.
On the other hand, nonlinear effects in~\eqref{Usoft} are pretty small, because
the exchanged graviton coupling $\ag$ affects only the $\qt_s$-dependence and
not the $\qt$-dependence. Therefore the single-exchange multiplicity
$\bk{N_1}\sim\bk{N}/\ag\sim\ord{\Theta_s^2}$ is down by a factor $\ag$ and
yields a quite limited enhancement, if any. By comparison, the nontrivial
feature of the state~\eqref{rcs} is that, though being confined to one emitted
graviton per exchange, it takes into account all exchanged graviton's
multiplicities and thus produces a reliable $\om R$ dependence.

To conclude, we stick in the following to the linear coherent state~\eqref{rcs}
to describe the main radiation features, but we introduce energy conservation
constraints also, to better understand the large $\om R$ part when approaching
the collapse regime.

%%%%%%%%%%%%%%%%%%%%%%%%%%%%%%%%%%%%%%%%%%%%%%%%%%%%%%%%%%%%%%%%%%%%%%%%%%%%%%%%%
\section{Finite angle radiation and approach-to-collapse\\ regime\label{s:far}}
%%%%%%%%%%%%%%%%%%%%%%%%%%%%%%%%%%%%%%%%%%%%%%%%%%%%%%%%%%%%%%%%%%%%%%%%%%%%%%%%%

%===============================================================================
\subsection{The emission amplitude in the sizeable angle region\label{s:rasr}}
%===============================================================================

In the following, we concentrate on the analysis of the amplitude~\eqref{ampRis}
in the semi-hard frequency region $\om R\gtrsim 1$, because the very soft
gravitons ($\om\ll b^{-1}$) are already well described by the approach of
sec.~\ref{s:gbsa}.

In that region the behaviour of~\eqref{ampRis} is quite sensitive to the angular
parameter $\Theta_E\equiv 2R/b$, which occurs in the amplitude in two ways: in
the overall coupling $\Theta_E\sqrt{\ag}$ and in the explicit expression for the
action, which is actually most sensitive, because of the $b=b_c$ branch-cut
(sec.~\ref{s:ram}). Note also the occurrence in the amplitude of
$\Delta(\bt-b\zt)$, which may be in the non-perturbative regime in the
integration region $|\bt-b\zt|\lesssim R$ in which its $S$-matrix factor may be
exponentially suppressed as in eq.~\eqref{elsup}.

For the above reason, we shall cutoff the rescattering contributions by the
requirement $|\bt-b\zt|>b_c$. If $\Delta(\bt)$ is in the perturbative
regime $\Theta_E\ll 1$, that change is subleading by a relative power of
$R^2/b^2$, because of phase space considerations, and the approach remains
perturbative. If instead $0<(b-b_c)/b\ll 1$, the cutoff procedure can be
extended to virtual corrections, by unitarizing the coherent-state operator, as
usual, but our approach becomes non-perturbative. Finally, we shall not discuss
at all --- in this paper --- the subcritical case $b\leq b_c$ by limiting
ourselves to the $b\to b_c^+$ approach-to-collapse regime. Considering $b < b_c$
would raise a variety of physical effects at both elastic and inelastic level
that deserve a separate investigation.

Since we limit ourselves to the $b\geq b_c$ case, we do not expect real problems
with $S$-matrix unitarity, because the tunneling suppression of the elastic
channel in eq.~\eqref{elsup} is absent. Nevertheless, the associated radiation
shows quite interesting features, especially in the approach-to-collapse regime
$b\to b_c^+$, that will be illustrated in the following.

Starting from the energy emission distribution of type~\eqref{spuv}
\begin{equation}\label{spruv}
  \frac{\dif E^\GW}{\dif\omega}
  = 2 Gs \frac{\Theta_E^2}{\pi^2} \int\frac{\dif^2\zt}{|z|^4}\;\left(
    \frac{\sin\omega R \Phi_{\ram,\cl}(\zt)}{\omega R}\right)^2
  \qquad (\om R\gtrsim 1)\;,
\end{equation}
we shall therefore distinguish two cases, in the large $\om R$ region:
\begin{itemize}
\item[a)] $\om
  R\gg\left(1-\frac{b_c^2}{b^2}\right)^{-3/2}\equiv(2\beta)^{-3/2}$.
  That is the truly small-angle regime, far away from the critical region
  $1-\frac{b_c^2}{b^2}\to 0$, or not too close to it. In that case $\om R$ is
  very large, which means very small $z$'s
  ($|y|^2\simeq|x|^2/\sqrt{\beta}\simeq\ord{1/\om R}$) so that the qualitative
  features of the radiation can be derived from the small-$z$ approximation of
  the modulating function
  \begin{equation}\label{PhiRcl}
    \Phi_{\ram,\cl} \simeq -\frac12\left[D_2(b) x^2 - D_1(b) y^2\right]+\ord{|z|^3} \;,
  \end{equation}
  where we have used the expansion
  \begin{subequations}
    \begin{align}
      \Phi_{\ram,\cl} &= \Delta(b)-\Delta(|\bt-b\zt|) - \Delta'(b)\bt\cdot\zt \label{PhiRamcl}\\
      &\simeq -\frac12 \frac{\partial^2\Delta}{\partial b_i\partial b_j} b^2 z_i
      z_j = \frac12\left[-\Delta''(b)b^2\hat{b}_i\hat{b}_j-b\Delta'(b)
        (\delta_{ij}-\hat{b}_i\hat{b}_j)\right] \label{Deltaexpn}
    \end{align}
  \end{subequations}
  yielding (by use of eqs.~\eqref{mc} and \eqref{osact})
  \begin{equation}\label{D12}
    D_1 = -b\Delta'(b) = \frac1{t_b} \;, \qquad
    D_2 = b^2\Delta''(b) = \frac{1+t_b^2}{t_b(3t_b^2-1)} \;.
  \end{equation}
  Here we note that $D_1\simeq D_2 \simeq 1$ for $b\gg b_c$, thus recovering
  eq.~\eqref{defI} discussed before, while $D_1\simeq\sqrt{3}$,
  $D_2\simeq\sqrt{2}\left(1-\frac{b_c^2}{b^2}\right)^{-1/2}$ for $\beta\ll 1$
  [eq.~\eqref{acrit}] and thus $D_2$ diverges for $\beta\to 0$. Correspondingly, we
  get a formula similar to~\eqref{spuv}:
  \begin{equation}\label{dEcl}
    \frac{\dif E_\cl^\GW}{\dif\om} = 2Gs\frac{\Theta_E^2}{\pi}
    \int\frac{\dif^2 z}{\pi|z|^4}\;\left[
      \frac{\sin\left(\frac{\om R}{2}(D_2 x^2-D_1 y^2)\right)}{\om R}\right]^2\;,
  \end{equation}
  where however we should assume $|x|^2/\sqrt{\beta}\ll|x|^{3/2}$ (or,
  $|x|\ll\beta$) whenever $\beta\ll 1$, because the actual behaviour of
  $\Delta(\bt-b\zt)$ is that of a branch cut with index $3/2$, with small
  convergence radius in the $x$-variable. We should therefore require $\om
  R\gg\beta^{-3/2}\gg 1$, as stated, so that such regime actually disappears in
  the limit $\beta\to 0$.
\item[b)] $1\ll \om R \ll \beta^{-3/2}$. That region opens up in the critical
  regime $0<\beta\ll 1$ and is dominant for $\beta\to 0$. However the quadratic
  small-$z$ expansion is no longer valid in the $x$-variable (because of the
  divergent coefficient), and the dominant approximation in the
  $|y|\simeq |x|\ll 1$ region becomes of type
  \begin{equation}\label{phiInter}
    -\Phi_{\ram,\cl}(z) \simeq -\Phi_1(x) \equiv 4\left[\frac13(\beta-x)^{3/2}
      -\frac13\beta^{3/2}+\frac12 x\sqrt{\beta}\right]
    \stackrel{|x|\ll \beta}{\simeq} \frac{x^2}{2\sqrt{\beta}} \;,
  \end{equation}
  where we have neglected, for simplicity, the $y$-dependence. We thus obtain
  what we shall call the 1-dimensional approximation to $\Phi_\ram$, which is
  easily derived by expanding all terms in the expression~\eqref{PhiRAM} of
  $\Phi_{\ram,\cl}$ for $0<\beta\ll 1$, both in $\beta$ and in $\zt$ and making
  the $(b-b_c)^{3/2}$ behaviour explicit by eq.~\eqref{acrit}.

  The striking feature of~\eqref{phiInter} is that, in the $\beta\ll|x|\ll 1$
  region, the dominant small-$x$ behaviour is $\Phi_1 \simeq (4/3)|x|^{3/2}$,
  reflecting the branch-cut of the ACV resummed action, which is responsible for
  the very large second derivative (large tidal force) in~\eqref{Deltaexpn}. By
  inserting that behaviour in~\eqref{spruv}, the corresponding distribution becomes
  \begin{equation}\label{enhancedSpectrum}
    \left.\frac1{\sqrt{s}}\frac{\dif E^\GW}{\dif\om}\right|_\text{enhanced}
    = \frac{\Theta_c^2}{\pi\om}\left(\frac{\om R}{3}\right)^{1/3}\Gamma(2/3)
    \;, \quad \Theta_c^2 \equiv \Theta_E^2(b_c) = \frac{8}{3\sqrt{3}}
  \end{equation}
  and falls off as $(\om R)^{-2/3}$ only. That radiation enhancement is a direct
  consequence of the critical index $3/2$ of the action branch cut at $b=b_c$.
\end{itemize}

We thus realize that, with increasing $R/b$, we quit the small-angle,
weak-coupling regime a) --- in which the radiated energy fraction is small (of
order $\Theta_E^2$) and shows at most a $\log(\om_M R)$ dependence with an
upper frequency cutoff $\om_M$ --- and we enter the strong-coupling regime b) in
which such fraction increases like $\Theta_E^2(\om_M R)^{1/3}$ thus endangering
the energy-conservation bound.

The possible violation of energy-conservation --- which is nevertheless taken
into account at linear level in the $\om$'s for rescattering --- is related to
the fact that the kinematical constraints are not explicitly incorporated in
multi-graviton production amplitudes and that multi-particle correlations are
neglected also. We shall introduce such constraints in sec.~\ref{s:ecc}.

%===============================================================================
\subsection{Radiation enhancement and scaling\label{s:rfgs}}
%===============================================================================

%-------------------------------------------------------------------------------
\subsubsection{Small-$\zt$ radiation spectrum}
%-------------------------------------------------------------------------------

In this section we present plots of the resummed amplitude and of the
corresponding radiated energy distribution obtained by numerical evaluation. In
this way we confirm the asymptotic behaviours derived in sec.~\ref{s:rasr} and
visualize the shape of such quantities in the transition regions.

\begin{figure}[ht]
  \centering
  \includegraphics[height=0.49\linewidth,angle=270]{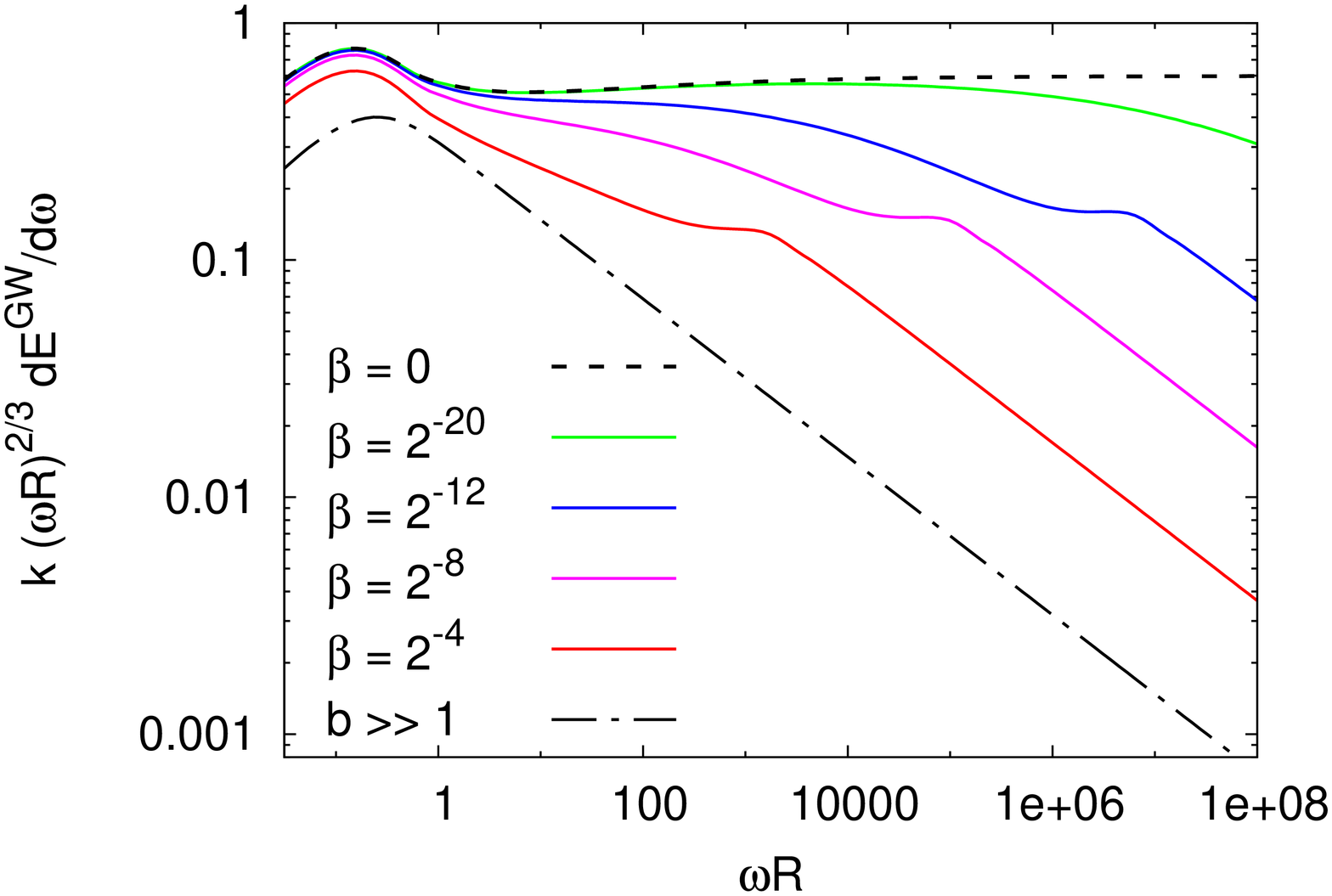}
  \includegraphics[height=0.49\linewidth,angle=270]{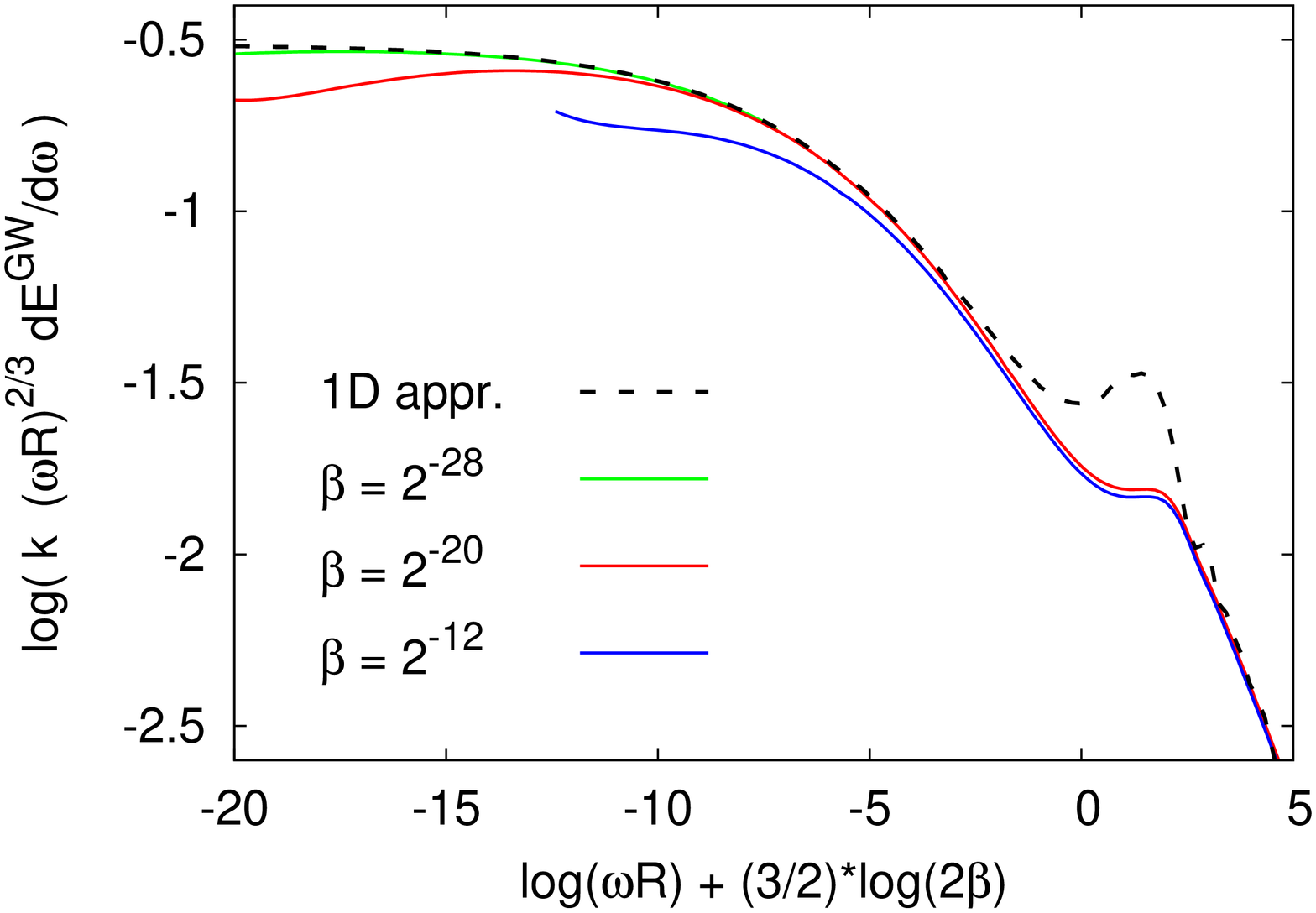}
  \caption{\it Left: the resummed spectrum for $\om R\gtrsim 1$ in the
    approach-to-collapse regime $b\to b_c$ ($\beta\to 0$) for various values of
    $\beta$ ranging from $1/2$ ($b\to\infty$) to 0 ($b\to b_c$). The spectrum
    has been multiplied by $(\om R)^{2/3}$ in order to highlight the enhancement
    in the intermediate regime $1\ll \om R \ll \beta^{-3/2}$, followed by the
    asymptotic $1/\om$ fall off. Right: some curves have been shifted
    horizontally and somewhat magnified in the neighbourhood of the transition
    region, showing the scaling behaviour with respect to the variable
    $a=(4\beta)^{3/2}\om R$; the black-dashed curve represents the
    one-dimensional approximate representation~\eqref{scaling}. As usual,
    spectra are reduced by the factor $k=(Gs\Theta_E^2)^{-1}$.}
  \label{f:resumPlot}
\end{figure}

Let us start by displaying the main features of the gravitational wave spectrum
obtained with the ACV resummation in the classical limit $\hom\ll E$ but close
to the collapse region $b\gtrsim b_c$. For $\om R\gtrsim 1$ this is obtained by
substituting the reduced-action model field~\eqref{PhiRAM} (actually its
classical limit $\Phi_{\ram,\cl}$ of eq.~\eqref{PhiRamcl}) in place of its
leading counterpart $\Phi$ inside eq.~\eqref{spuv}. The results for various
values of $\beta$ are shown in fig.~\ref{f:resumPlot}. According to the
estimates in sec.~\ref{s:rasr}, at smaller and smaller $\beta\ll 1$ there is a
larger and larger intermediate region $1\ll\om R\ll \beta^{-3/2}$ of reduced
decrease of the frequency spectrum $\sim\omega^{-2/3}$, followed by the typical
asymptotic $\om^{-1}$ fall off at $\om R\gg \beta^{-3/2}$. In order to better
discriminate the two regimes, the spectrum has been multiplied by $(\om
R)^{2/3}$, so that in the intermediate enhanced region the curves are almost
flat.

In the first plot of fig.~\ref{f:resumPlot} the black dot-dashed curve ($\beta=1/2$)
represents the small-angle spectrum described in sec.~\ref{s:lorea}. Decreasing
the value of $\beta$ we obtain the solid curves (red, magenta, blue, green) and
we observe the expected enhancement that amounts to a numerical factor of order
one for $\om R\lesssim 1$, but becomes much more important for large
$\om R\gtrsim 1$. It is also clear that the extension of the enhanced regime
increases while decreasing $\beta$. In the limit $\beta\to 0$ the rescaled
spectrum approaches the almost horizontal dashed line.

It is apparent that the shapes of the curves are quite similar at large $\om R$,
including the transition region between the enhanced and asymptotic regimes. By
rescaling the independent variable $\om R\to\om R \beta^{3/2}$, the curves at
small $\beta$ go on top of each other, as shown in the right plot of
fig.~\ref{f:resumPlot}. In other words, the asymptotic shape of the spectrum is
a function of the single variable $a\equiv (4\beta)^{3/2}\om R$.  This scaling
property can be understood by exploiting the small-$z$
expansion~\eqref{phiInter} that, substituted into eq.~\eqref{spuv}, provides the
approximate representation
\begin{align}
  \frac{1}{Gs\Theta_E^2}(\om R)^{2/3} \frac{\dif E}{\dif\om} &\simeq
  \frac{2}{\pi} a^{-1/3} \int_{-1}^{\infty} \dif t \;
  \frac{\sin\left\{a\left[\frac13\left((1+t)^{3/2}-1\right)-\frac{t}{2}\right]\right\}}{
    |t|(1+\sqrt{1+t})} \xrightarrow{a\to 0} \frac{2\Gamma(2/3)}{3^{1/3}\pi} \nonumber \\
  a &\equiv (4\beta)^{3/2}  \om R \label{scaling}
\end{align}
which depends only on the scaling variable $a$. This function is displayed in
the black dashed line on the right plot of fig.~\ref{f:resumPlot}, and it describes
well the scaling behaviour in the enhanced region $a\ll 1$, and reasonably well
the large-$\om R$ region $a\gg 1$.

%-------------------------------------------------------------------------------
\subsubsection{Angular behaviour\label{s:ab}}
%-------------------------------------------------------------------------------

\begin{figure}[ht]
  \centering
  \raisebox{28mm}[0mm][0mm]{$A_x$}%
  \includegraphics[width=0.46\linewidth]{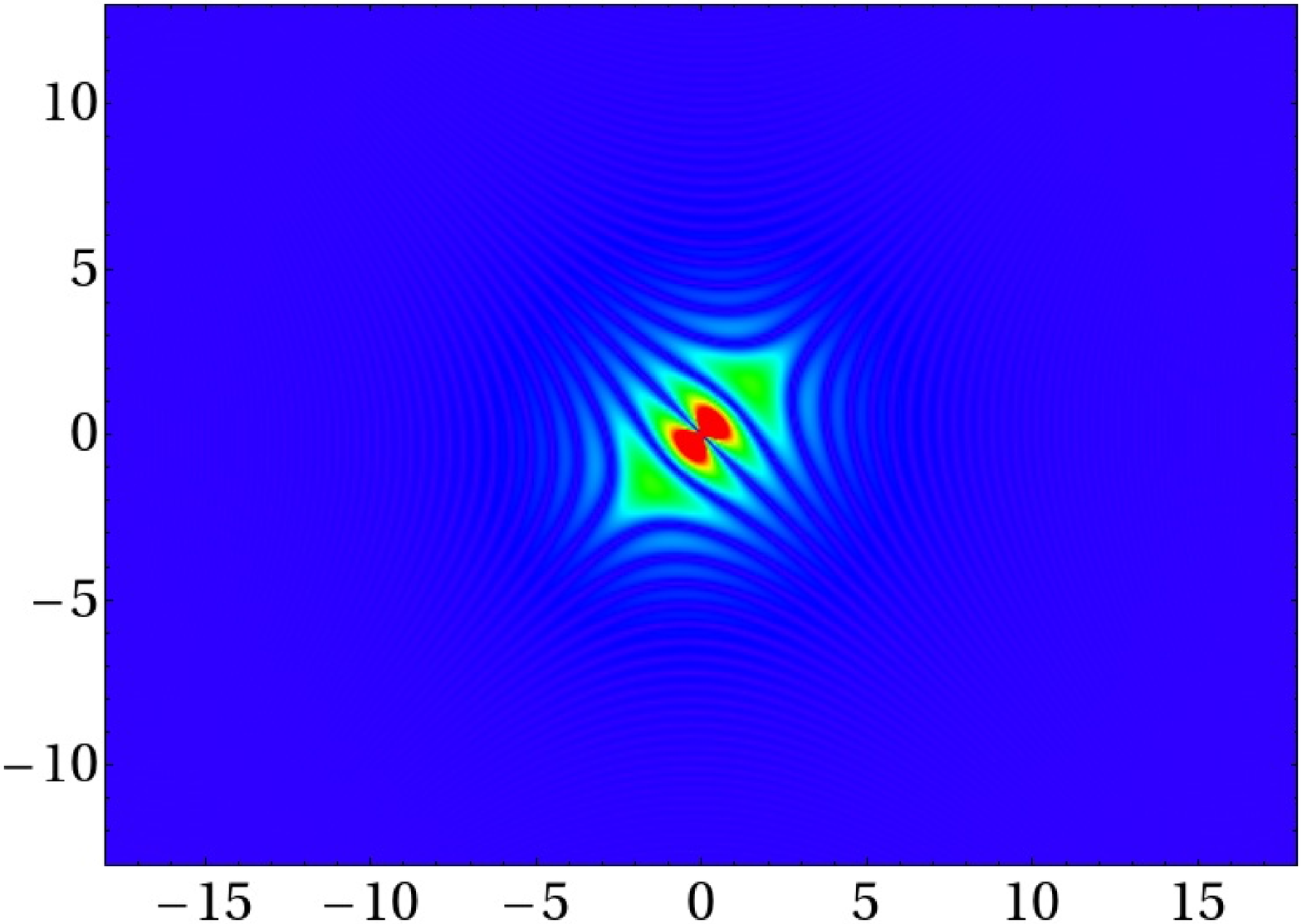}
  \hspace{0.02\linewidth}
  \includegraphics[width=0.46\linewidth]{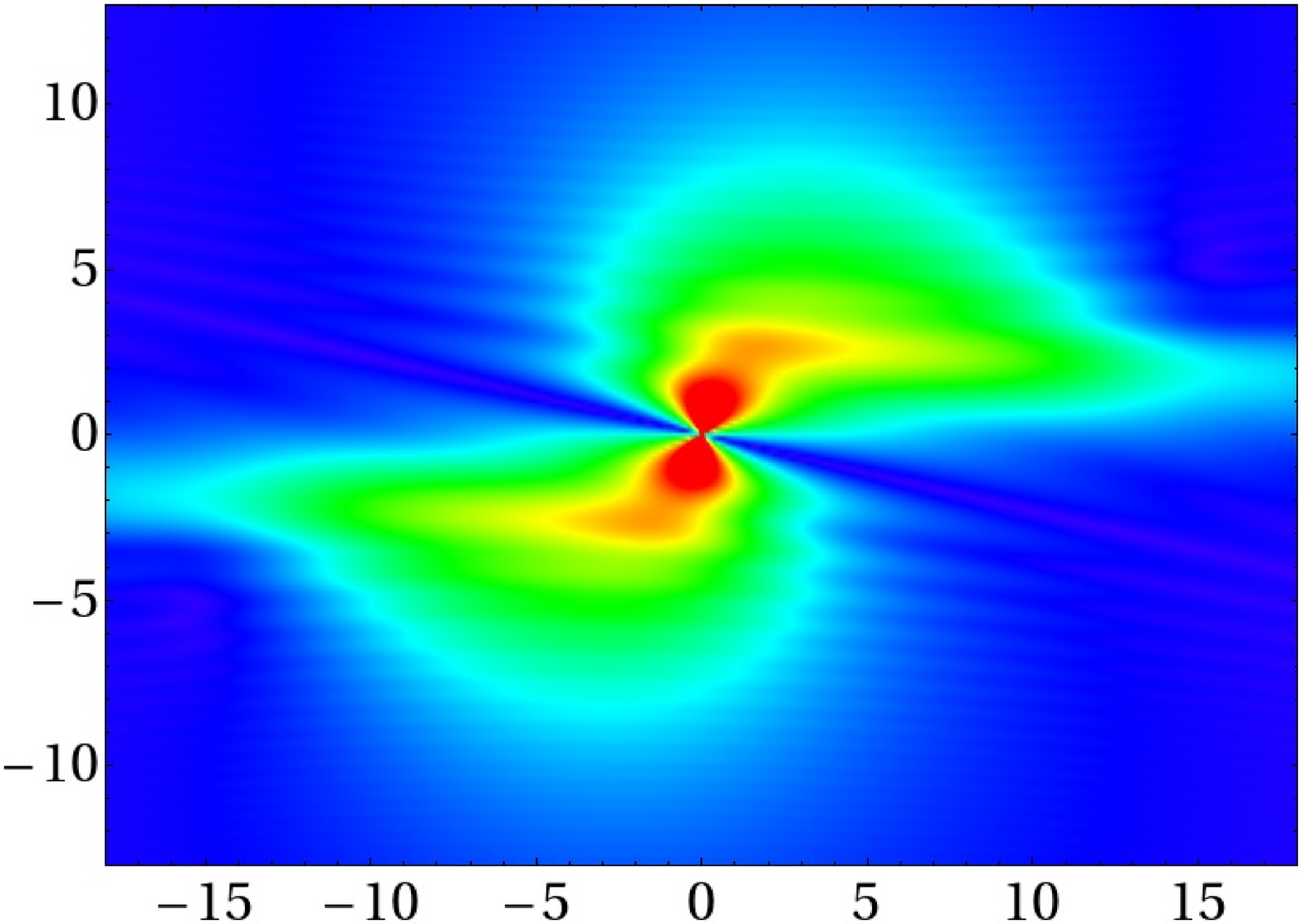}\\
  $A_y$\hspace{50mm}\null\\
  \hspace{0.06\linewidth} (a) \hspace{0.45\linewidth} (b) \\
  \caption{\it Emission pattern of gravitational radiation for $\lambda=-2$ on
    the tangent space centered at $\Tht_s$, parametrized by
    $\At\equiv \sqrt{\om R}\, \frac{\tht-\Tht_s}{|\Tht_s|}$. {\rm(a)} the
    ``isotropic'' case $D_1=D_2$ discussed in sec.~\ref{s:lorea}; {\rm(b)}
    ``anisotropic'' case with $D_1=\sqrt{3}$, $D_2=32$.}
  \label{f:azimQuadratic}
\end{figure}

The angular behaviour of graviton radiation associated to large scattering
angles $\Theta_s\sim 1$ can be obtained by numerical integration of the
amplitude~\eqref{ampRis}. However, in the main region of the spectrum, namely
$\om R\gtrsim 1$, it can be more conveniently described by using the small-$z$
approximation~\eqref{PhiRcl} of the modulating function $\Phi_{\ram,\cl}$. The
main point here is that the two dispersion coefficients $D_1$ and $D_2$, which
are equal for small scattering angle, become more and more different when
approaching the critical angle $\Theta_c$. This fact causes the ensuing
distribution of graviton radiation to be more and more directional, still
concentrated at $\tht\simeq\Tht_s$, but with a larger dispersion, in particular
along the $x$-direction, i.e., that of the scattering plane. This is clearly
seen in fig.~\ref{f:azimQuadratic}, where we compare on the
$\At\equiv \sqrt{\om R}\, \frac{\tht-\Tht_s}{|\Tht_s|}$ plane the ``isotropic''
radiation (a) when $D_1=D_2=1$ with the ``anisotropic'' case (b) $D_1=\sqrt{3}$,
$D_2=32$ (corresponding to $\beta\simeq 0.001 $).

We see, first of all, that in the collapse region (fig.~\ref{f:azimQuadratic}b)
the radiation is strongly enhanced, still keeping its correlation with the
outgoing particle 1' in the overall picture of the two jets. Furthermore, the
larger dispersion in $\theta_x$ compared to $\theta_y$ gives a rationale for the
1-dimensional approximation~\eqref{scaling} in the conjugated variables $x$ and
$y$. Finally, such features are valid for any given frequency range $\Delta\om$
and are thus somewhat independent of their relative normalization, which is
possibly affected by energy-conservation constraints, to be discussed next.

%===============================================================================
\subsection{Energy-conservation and ``temperature''\label{s:ecc}}
%===============================================================================

In order to take into account energy-conservation constraints, we shall
calculate coherent-state amplitudes and distributions by setting --- event by event --- the
explicit energy bound $\sum_{i=1}^N\hom_i<E$, in which we refer to a single
``jet'', say along $p_1$.%
\footnote{The point is that the energy of the forward (backward) gravitons is
  essentially taken at the expenses of the sole particle 1 (2).}
Such bounds are effectively extended to virtual corrections by a factorization
assumption, as proposed by~\cite{GriVe_pc} on the basis of the
AGK~\cite{AGK72} cutting rules (see also sec.~4.3 of ref.~\cite{ACV88}).

More explicitly, we modify the original independent-particle
distributions~\eqref{genFun} in a radiation sample of energy up to
$E$ by introducing the corresponding kinematical bounds together with a
rescaling factor $1/N(E)$ in probability (or $1/\sqrt{N(E)}$ in amplitude) to be
determined by unitarity. For instance, by considering for simplicity the
$\om$-variables only, we define the energy-conserving distributions
\begin{equation}\label{partDist}
  \tilde{P}_0=\frac{P_0}{N(E)} \;, \qquad \dif\tilde{P}(\{\om_i N_i\}) =
  \frac{P_0}{N(E)}\prod_{i} \frac{[p(\om_i)\Delta(\om_i)]^{N_i}}{N_i!}
  \Theta\Big(E-\sum_{i} \hom_i N_i\Big)\;,
\end{equation}
where the $p(\om)$ density is given by~\eqref{enDist}, with the amplitude
$\ampRid_\lambda(\bt,\vq)$ in~\eqref{ampRis}. We have also discretized the Fock
space in regions of extension $\Delta(\om_i)$, containing a number of gravitons
$N_i$ each.

The normalization factor $N(E)>0$ in~\eqref{partDist} is determined by the
unitarity condition $\sum_{\{N_i\}} \tilde{P}(\{N_i\}) = 1$ and takes the form
[cf.~eq.~\eqref{genFun}]
\begin{equation}\label{NE}
  N(E) = \int_{-\ui\infty}^{+\ui\infty}\frac{\dif\lambda}{2\pi\ui}\;
  \frac{\esp{\lambda E}}{\lambda+\e}\exp\left\{\int_0^\infty\dif\om\; p(\om)
  \left[\esp{-\om\lambda}-1\right]\right\}
\end{equation}
which carries the energy-conservation constraints and is obtained by summing
over all events the (positive) partial probabilities. We stress the point that
$E$ in eq.~\eqref{NE} is the energy available for the measures being considered, so that
$E=\sqrt{s}/2$ if we consider the whole jet, but becomes $\sqrt{s}/2-\hbar\om$
if we consider events associated to an observed graviton $\om$ in that jet, and
so on. On the basis of eqs.~\eqref{partDist} and~\eqref{NE} it is
straightforward to obtain, for the inclusive distributions,
\begin{equation}\label{incDist}
  \frac{\dif\Num}{\dif\om} = p(\om)\frac{N(E-\hom)}{N(E)} \;, \qquad
  \frac{\dif^2\Num}{\dif\om_1\dif\om_2} = p(\om_1)p(\om_2)\frac{N(E-\hom_1-\hom_2)}{N(E)}
\end{equation}
and so on. We notice also that virtual corrections are explicitly incorporated
in~\eqref{NE} via the normal-ordering of the state~\eqref{rcs} [cf.\
eq.~\eqref{P0}], and that $N(E)$ is actually infrared safe.

The main point is now that the inclusive distributions~\eqref{incDist} carry
$N(E)$-dependent correction factors due to the phase-space restrictions
$E\to E-\hom,\cdots$, and so on, that will turn out to suppress the
large-$\om R$ region by an exponential cutoff. Arguments for a cutoff are
provided also in the approach of ref.~\cite{Addazi:2016ksu} to the
transplanckian scattering without impact parameter identification of
ref.~\cite{Dvali:2014ila}.

In order to estimate $N(E)$ it is convenient to rewrite it in terms of the
quantity ($\lambda\equiv R\tau$)
\begin{equation}\label{vmOmE}
  \frac{\bk{\hom}_\tau}{\sqrt{s}/2} \equiv F(\tau) = \int_0^\infty\dif\om\;
  \frac{\hom}{\sqrt{s}/2} p(\om) \esp{-\om R\tau} \;,
\end{equation}
which represents the (exponentially weighted) radiated-energy fraction, given in
our case \eqref{ampRis} by [cf.~eq.~\eqref{spruv}]
\begin{equation}\label{Ftau}
  F(\tau) = \frac{\Theta_E^2}{\pi^2}\int\frac{\dif^2 z}{|z|^4}
  \int_0^\infty \frac{\sin^2(\om R\Phi_{\ram,\cl})}{(\om R)^2}\esp{-\om R\tau}
  \dif(\om R) \;.
\end{equation}
We then obtain from eq.~\eqref{NE} ($\ag=R\sqrt{s}/2=Gs$) the expression
\begin{equation}\label{NEint}
  N(E) = \text{const}\int_{\e-\ui\infty}^{\e+\ui\infty}\dif\tau\; \exp\left\{
    ER\tau-\log\tau-\ag\int_0^\tau F(\tau')\;\dif\tau'\right\}
\end{equation}
and we proceed to estimate it by the saddle-point method. The saddle point value
$\tb>0$ is determined by the equation ($E=\sqrt{s}/2$)
\begin{equation}\label{speq}
  F(\tb) = 1-\frac1{\ag\tb} \;,
\end{equation}
which represents the share between emitted (l.h.s.) and preserved
($1/\ag\tb$) energy fractions at the saddle point exponent
$\tb$. Fluctuation corrections are also calculable (app.~\ref{a:fc}) and will be
discussed shortly.

\begin{figure}[ht]
  \centering
  \includegraphics[height=0.49\linewidth,angle=270]{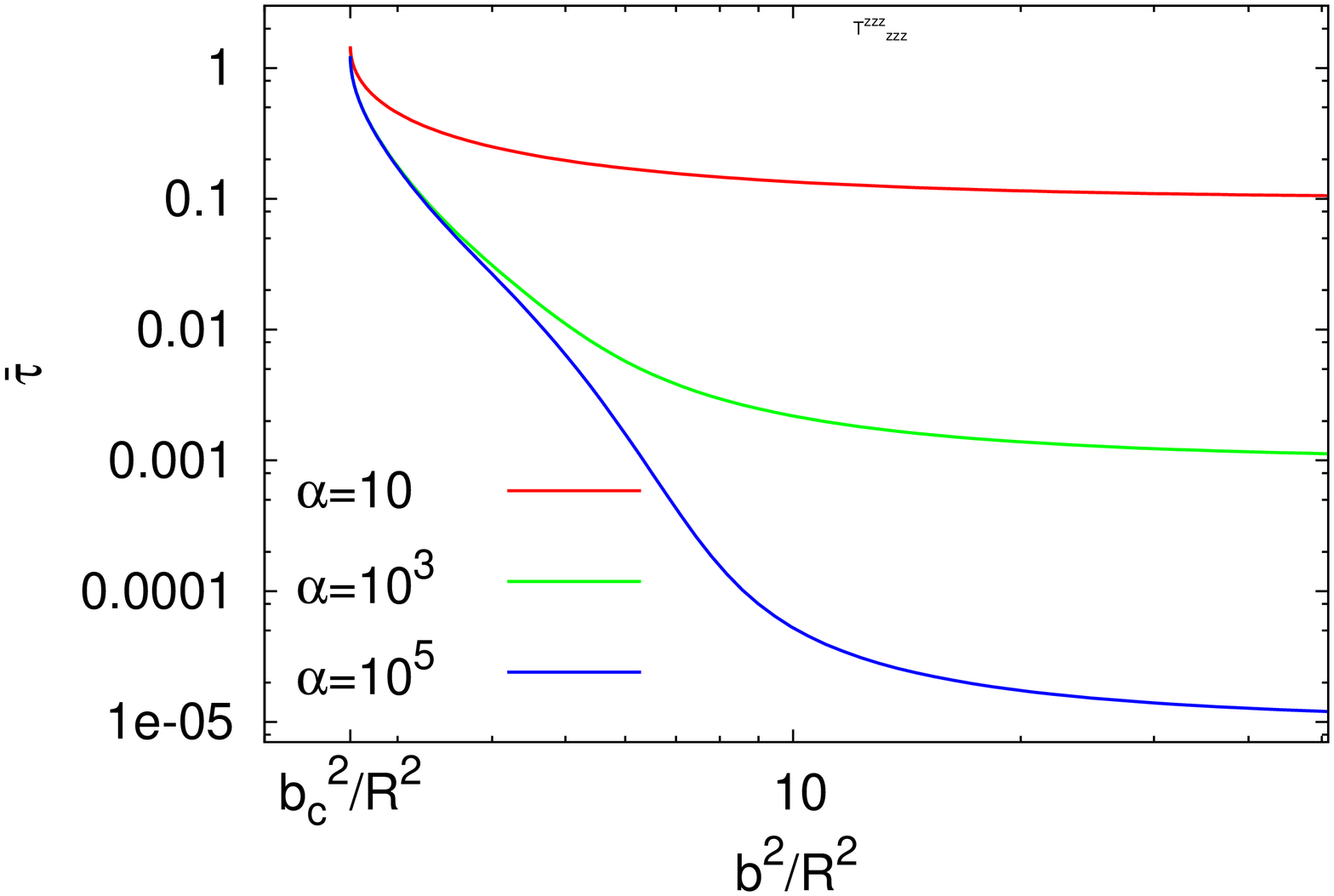}
  \includegraphics[height=0.49\linewidth,angle=270]{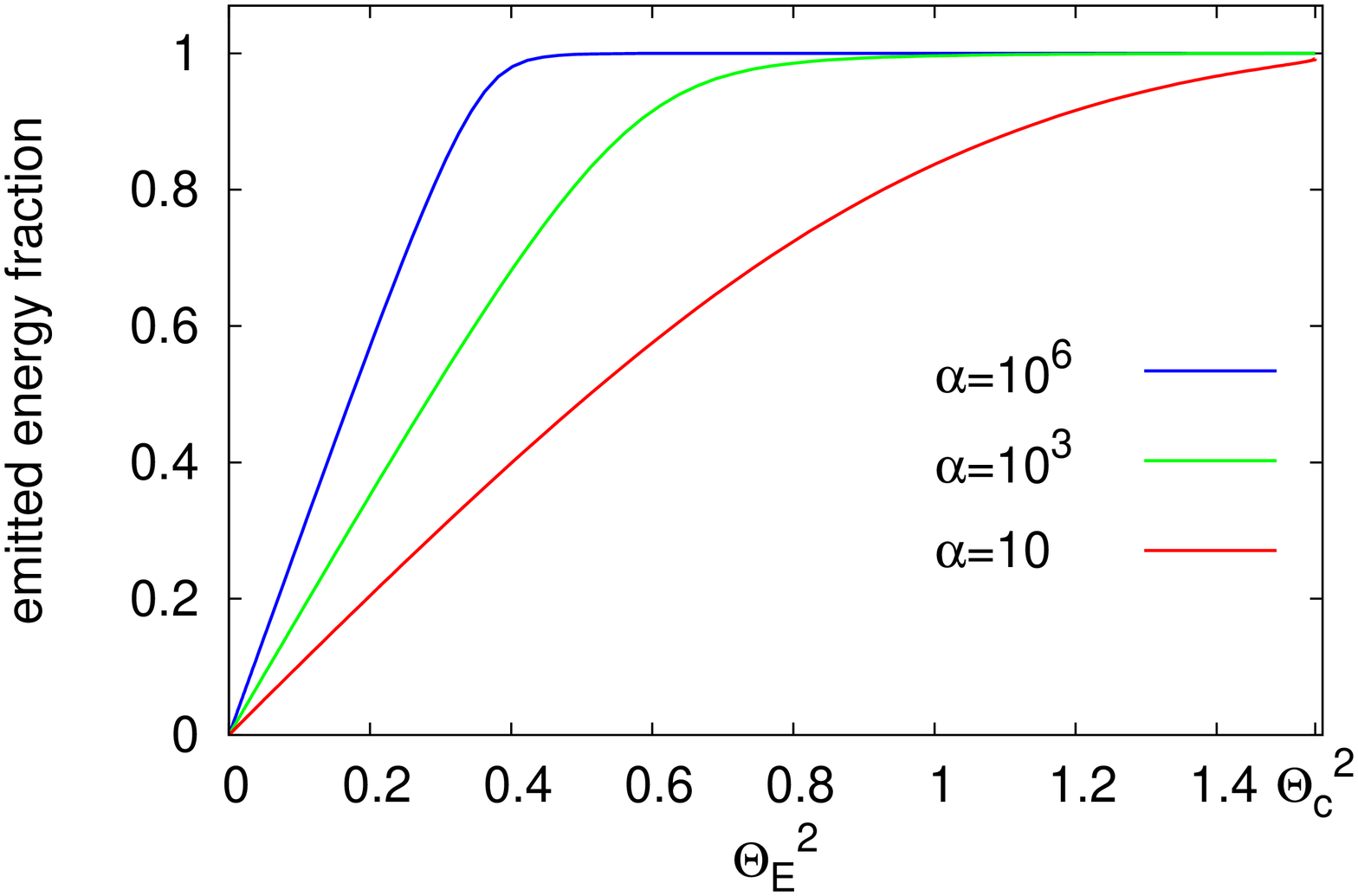}\\
  \hspace{0.03\linewidth} (a) \hspace{0.48\linewidth} (b) \\
  \caption{\it {\rm(a)} Dependence of the saddle point $\tb$ on the impact parameter
    $b^2$, for various values of $\ag$; {\rm(b)} emitted energy fraction vs
    $\Theta_E^2$ for various values of $\ag$.}
  \label{f:sp}
\end{figure}

The numerical evaluation of~\eqref{speq} (fig.~\ref{f:sp}) is better understood
by working out eq.~\eqref{Ftau} in the form
\begin{equation}\label{sp2}
  1-\frac1{\ag\tb} = F(\tb) = \frac{\Theta_E^2}{\pi^2}
  \int\frac{\dif^2\zt}{|z|^4}\; |\Phi(z)| I\left(\frac{2|\Phi(\zt)|}{\tb}\right)
  \;,
\end{equation}
where, by explicit integration,
\begin{equation}\label{tgchi}
  \tan\chi I(\tan\chi) = \chi\tan(\chi) + \frac12 \log(\cos^2\chi) \;,
\end{equation}
The result shows that $\tb\sim\ord{1/\ag}$ in the small-angle region ($b\gg R$),
while $\tb=\ord{1}$ in the collapse regime. In between, the radiated energy
fraction varies from 0 to 1.

In order to understand the role of $\tb$ for the energy-conservation cutoff, we
estimate the inclusive distribution~\eqref{incDist} at the saddle point, and we
find
\begin{equation}\label{dNdom}
  \frac{\dif\Num}{\dif\om} = p(\om)\esp{-(\tb+\Delta\tau)\om R} \;,
\end{equation}
where the $\tb$ term in the exponent comes from the explicit energy dependence
of $N(E-\om)$, and the correction $\Delta\tau$ comes from the implicit one
through $\tb(E-\om)$, to which --- by $\tb$-stationarity --- mostly fluctuations
contribute. We show in app.~\ref{a:fc} that this kind of corrections is sizeable
when $\tb=\ord{1/\ag}$ is small (where however the cutoff is not really
important) while it is small when $\tb=\ord{1}$ is essential, that is in the
approach-to-collapse regime. The cutoff exponent $\tb$ has already been used in
the definitions~\eqref{Ftau} and~\eqref{sp2}.

\begin{figure}[ht]
  \centering
  \includegraphics[width=0.49\linewidth]{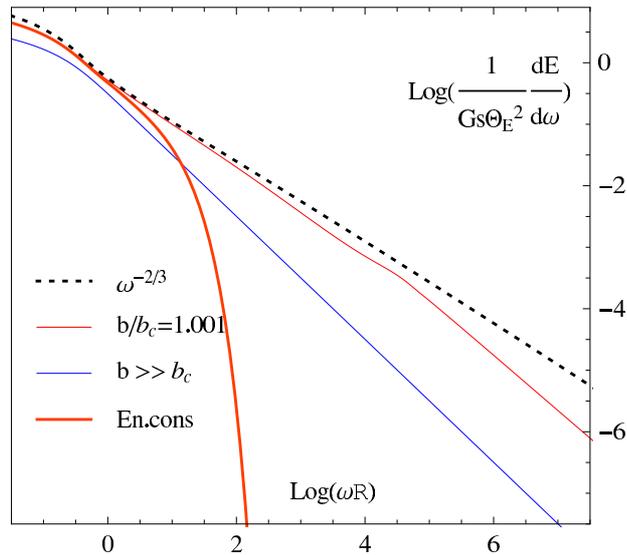}
  \caption{\it Frequency spectrum of graviton radiation in various
    contexts. Three curves have no energy conservation constraints, and
    correspond to:
    large $b/R$ where subleading effects are negligible (blue);
    $b$ close to the critical parameter $b_c$
    where subleading effects causes the enhancement (thin red);
    limit $b\to b_c^+$ (dashed black). The last curve (thick red) shows the
    suppression due to energy conservation constraints in the case of $b$ close
    to $b_c$.}
  \label{f:spettroCutoff}
\end{figure}

In more detail, it is useful to distinguish a very small angle regime
$\Theta_s^2\ll\bar\Theta^2 \equiv 1/\log\ag$, in which $\bar\Theta^2$ acts as
threshold for important energy-conservation effects like energy fractions of
order 0.5, say [fig.~\ref{f:sp}.b and eq.~\eqref{kinBound}]. Below it, the
radiated fraction $\sim\Theta_s^2$ is very small and so is $\tb\sim
1/\ag$. Furthermore, the exponent $\tb+\Delta\tau\sim\Theta_s^2/\ag$ is even
smaller than $\tb$ because of cancellations with the term $\Delta\tau$
(app.~\ref{a:fc}), thus leading to negligible conservation corrections.

On the other hand, for $\Theta_s^2$ above $\bar\Theta^2$, both the exponent part
$\tb$ and the radiated fraction increase (fig.~\ref{f:sp}) up to $\ord{1}$ for
$\Theta_E^2\to\Theta_c^2 = \ord{1}$, while $\Delta\tau/\tb$ becomes
$\ord{1/\ag}\ll 1$, that is, small.

In that case --- of strong coupling and radiation enhancement --- the whole
energy is radiated off, and this fact fixes $\tb=\tau_c=1.2$ in a rather precise
way.  Furthermore, the same exponent (with $\Delta\tau/\tb\sim 1/\ag\ll 1$)
occurs in all the graviton distributions~\eqref{incDist} which --- because of
such approximate universality --- turn out to be approximately factorized and
thus weakly correlated, even after the inclusion of energy conservation.  In
other words, while the rescaling factor $\sqrt{N(E)}$ keeps the phase relations
of the coherent state~\eqref{rcs} among the various $\om$-bins, it also
introduces, by the $E$-dependence of~\eqref{NE} and the $\om$-dependence
of~\eqref{incDist}, an almost universal frequency cutoff parameter $R^{-1}$, a
``quasi-temperature'' we would say, in the approach-to-collapse regime.
Numerically, the exponent $\tau_c R$ turns out to be of the order of the inverse
Hawking temperature for a black hole mass $\simeq 0.1 \sqrt{s}$, notably smaller
than $\sqrt{s}$, and the corresponding spectrum --- in each one of the two jets
with $E=\sqrt{s}/2$ that our radiation consists of --- is given in
fig.~\ref{f:spettroCutoff}.

Our semiclassical method does not allow, at present, a precise interpretation of
the features just mentioned in terms of black hole physics, mostly because of
our ignorance of what a black hole really is in quantum physics. Nevertheless we
think that, applying our soft-based representation to the approach-to-collapse
regime, we have constructed a coherent radiation sample which shares some of its
properties with a Hawking radiation, thus suggesting a deeper relationship. That
fact, because of coherence, goes in the direction of a quantum theory overcoming
the information paradox, even if the details of such relationship are not known
yet.

%%%%%%%%%%%%%%%%%%%%%%%%%%%%%%%%%%%%%%%%%%%%%%%%%%%%%%%%%%%%%%%%%%%%%%%%%%%%%%%%
\section{Outlook\label{s:o}}
%%%%%%%%%%%%%%%%%%%%%%%%%%%%%%%%%%%%%%%%%%%%%%%%%%%%%%%%%%%%%%%%%%%%%%%%%%%%%%%%

The main technical progress presented here is the extension of the semiclassical
graviton radiation treatment in transplanckian scattering to cover finite
scattering angles $\sim R/b$. That result is in turn based on the ACV eikonal
resummation and on the validity --- for such reduced-action model --- of the
soft-based representation of the radiation amplitude argued for in
sec.~\ref{s:rm}.

After such steps, we are really able to follow the approach to the classical
collapse regime by a fully explicit, unitary coherent state, given the fact that
collapse is signalled by a branch cut singularity of the action at
$b=b_c=\ord{R}$ with some scattering angle $\Theta_c=\ord{1}$ and branch-cut
index $3/2$. While $b_c$ and $\Theta_c$ are expected to be somewhat
model-dependent, the index $3/2$ is expected to be robust because it yields the
first non-analytic behaviour, by the action stationarity in the angular
parameter $t_b$.

The first striking feature that we notice is that, because of the index $3/2$,
the action has very large second derivatives (tidal forces) and thus yields a
radiation enhancement causing about the whole-energy be radiated off for
$b\to b_c^+$. Actually, it also requires the enforcement of the kinematical
constraints in order to insure energy conservation.

Energy-conservation constraints (sec.~\ref{s:far}) are introduced in real
emission event by event and transferred to virtual corrections in some
approximation which amounts to a factorization assumption, natural for the
weakly correlated coherent-state that we have constructed. The outcome is that
energy-conservation effects, which are negligible for $\Theta_s\ll 1$, are
instead quite important in the approach-to-collapse regime, and provide an
exponential suppression of the large $\om R$ region. The latter is approximately
universal, that is occurs in all the inclusive distributions, with small
corrections and weak correlations, both depending on the parameter $1/\ag$,
where $\ag=Gs/\hbar\gg 1$ is the magnitude of the final multiplicity.

The conclusive features just mentioned show that our radiation sample
(corresponding to two jets with masses up to $\sqrt{s}/2$) --- though coherent
by construction --- is characterized by an almost universal, exponential
frequency cutoff close to $1/R$ which plays a role analogous to the Hawking
temperature (at a mass notably smaller than $\sqrt{s}$). Such fact suggests a
deeper relationship with the possible collapse dynamics, whose boundaries
are however difficult to pinpoint, in view of both our approximations and our
ignorance about the nature of a quantum black hole. We nevertheless think ,
because of coherence, that our results go in the direction of a quantum theory
overcoming the information paradox, even if details of their relationship to
black hole physics are not known yet.

\section{Acknowledgments}

It is a pleasure to thank Gabriele Veneziano for a number of interesting
conversations on the topics presented in this paper, and Domenico Seminara for
useful discussions. We also wish to thank the {\em Galileo Galilei Institute for
  Theoretical Physics} for hospitality while part of this work was being done.

%%%%%%%%%%%%%%%%%%%%%%%%%%%%%%%%%%%%%%%%%%%%%%%%%%%%%%%%%%%%%%%%%%%%%%%%%%%%%%%%%
%%%%%%%%%%%%%%%%%%%%%%%%%%%%%%%%%%%%%%%%%%%%%%%%%%%%%%%%%%%%%%%%%%%%%%%%%%%%%%%%%
\appendix
%%%%%%%%%%%%%%%%%%%%%%%%%%%%%%%%%%%%%%%%%%%%%%%%%%%%%%%%%%%%%%%%%%%%%%%%%%%%%%%%%
%%%%%%%%%%%%%%%%%%%%%%%%%%%%%%%%%%%%%%%%%%%%%%%%%%%%%%%%%%%%%%%%%%%%%%%%%%%%%%%%%
\section*{Appendices}

%%%%%%%%%%%%%%%%%%%%%%%%%%%%%%%%%%%%%%%%%%%%%%%%%%%%%%%%%%%%%%%%%%%%%%%%%%%%%%%%
\section{Fluctuation corrections to inclusive distributions\label{a:fc}}
%%%%%%%%%%%%%%%%%%%%%%%%%%%%%%%%%%%%%%%%%%%%%%%%%%%%%%%%%%%%%%%%%%%%%%%%%%%%%%%%

It is straightforward to introduce a quadratic fluctuation expansion in
eq.~\eqref{NEint} to yield the normalization factor
\begin{align}
  N(E) &= \int_{\tb-\ui\infty}^{\tb+\ui\infty}\frac{\dif\tau}{2\pi\ui}\;\exp\left\{
    ER\tb-\log\tb-\ag\int_0^{\tb} F(\tau')\dif\tau' +\frac12 (\tau-\tb)^2
    \left[\frac1{\tb^2}-\ag F'(\tb)\right]\right\} \nonumber \\
 &\simeq \text{const}\exp\left\{\left[ER\tb-\ag\int_0^{\tb} F(\tau')\dif\tau'
     \right]-\frac12\log\left(1-\ag\tb^2 F'(\tb)\right)\right\} \;.\label{fluc}
\end{align}
We note that the $\log\tb$ term cancels out, so that the overall size of
fluctuations is determined by the function
\begin{equation}\label{ftau}
  f(\tb) \equiv -\ag\tb^2 F'(\tb) \simeq
  \begin{cases}
   \displaystyle \frac{\Theta_E^2}{2\pi}\ag\tb & (\Theta_E\ll 1) \\[4mm]
   \displaystyle \frac13\ag\tb^{2/3}\tau_c^{1/3} &
   (\Theta_E\simeq\Theta_c) \;.
  \end{cases}
\end{equation}
Although this function is pretty small (large) in the small (large) angle
regime, its relative importance with respect to the exponent part $\tb$ goes
just in the opposite. In fact, in the small-$\tb$ regime (where the cutoff is
unimportant) the expansion of the remaining log term produces contributions of
order comparable to those in square brackets.

To better understand this point, we combine eq.~\eqref{fluc} with the
saddle-point equations
\begin{equation}\label{spe}
  ER = \ag F(\tb)+\frac1{\tb} \;, \qquad \tb'(E) = -R\tb^2 \frac1{1+f(\tb)}
\end{equation}
to get, after some algebra,
\begin{align}
  \left.\frac{\dif\log N(E-\om)}{\dif E}\right|_{\om=0} &= R\tb(E)
  +\left.\frac{\partial\log N}{\partial\tb}\right|_{\tb(E)}\tb'(E)
  \nonumber \\
  &= R\tb(E) \left[\frac{f(\tb)}{1+f(\tb)} +\frac{\tb f'(\tb)}{2(1+f(\tb))^2}
    \right] = R \tb \left(1+\frac{\tb}{2}\frac{\partial}{\partial\tb}\right)
    \left(\frac{f}{1+f}\right) \;.
\end{align}
Consider first the strong coupling regime in which $\tb=\ord{1}$. It is clear
that $f(\tb)$ is $\ord{\ag}$ so that
\begin{equation}\label{dNdomega}
  \frac{\dif\Num}{\dif\om} \simeq p(\om) \esp{-\om R(\tb+\Delta\tau)}
  \simeq p(\om)\esp{-\om R\tb}
\end{equation}
with a correction
\begin{equation}\label{Deltatau}
  \frac{\Delta\tau}{\tb} = \left( 1+\frac{\tb}{2}\frac{\partial}{\partial\tb}\right)
  \left(\frac{-1}{1+f}\right)
\end{equation}
which is small, $\Delta\tau/\tb=\ord{1/\ag}$, leading to an approximately
universal exponent $\tb$.

On the other hand, in the weak coupling regime, $\Theta_E\ll 1$, $\tb$ is small,
starting $\ord{1/\ag}$ so that $f=\ord{\Theta_E^2}$ is small too. As a
consequence, relative corrections are large, so as to allow cancellations with
the leading term and an even smaller exponent. That is fortunately unimportant,
because energy conservation corrections are small in that regime. For instance,
in the regime $\Theta_E^2 \ll 1$ and $\tb=\ord{1/\ag}$, we get
\begin{equation}\label{tau+dt}
  \tb+\Delta\tau \simeq \tb\frac{f}{1+f} = \ord{\Theta_E^2/\ag} \;,
\end{equation}
yielding negligible corrections to the naive inclusive distribution. We conclude
that for $\Theta_E\ll 1$ inclusive distributions avoid the energy-conservation
cutoff, while for $\Theta_E\simeq\Theta_c=\ord{1}$ such cutoff is provided by
$\tb$ and is approximately universal. The final multiplicity is provided
by~\eqref{dNdomega} and is of $\ord{\ag}$ with a finite coefficient.

%%%%%%%%%%%%%%%%%%%%%%%%%%%%%%%%%%%%%%%%%%%%%%%%%%%%%%%%%%%%%%%%%%%%%%%%%%%%%%%%
\section{One-dimensional integral representation of the amplitude at large
  $\bs{\om R}$\label{a:angular}}
%%%%%%%%%%%%%%%%%%%%%%%%%%%%%%%%%%%%%%%%%%%%%%%%%%%%%%%%%%%%%%%%%%%%%%%%%%%%%%%%

We want to give a simple representation of the graviton emission amplitude for
large $\om R \gg 1$. According to the discussion in sec.~\ref{s:lorea}, the
emission amplitude $\ampRid$ is dominated by the small-$z$ region, where the
modulation function $\Phi_R$ can be approximated by its quadratic
expansion~\eqref{PhiRcl}.  We can therefore express $\ampRid$ in terms of the
two-dimensional complex integral
\begin{equation}\label{igD}
  I(\At)=\int\frac{\dif^2 Z}{2\pi\ui}\;\frac{\esp{\ui(Z A^* +Z^* A)}}{Z^{*2}}
  \left[ \esp{\ui(D_2 x^2-D_1 y^2)}-1 \right]
   \;, \qquad(Z=x+\ui y) \;,
\end{equation}
as in eq.~\eqref{defI}. For the resummed amplitude we note the presence of the
two dispersion coefficients $D_1$ and $D_2$ that are different for finite $b$.

Our aim here then is to provide a simple representation of $I(\At)$.  Gaussian
integration is possible by eliminating the double pole by derivation with
respect to $A$:
\begin{align}
  -\frac{\partial^2}{\partial A^2}I(A,A^*) &=
  \int\frac{\dif x\,\dif y}{2\pi\ui} \; \esp{\ui(2 x A_1 + 2 y A_2)}
  \left[\esp{\ui(D_2 x^2-D_1 y^2)}-1\right] \nonumber \\
  &= \frac1{2\ui\sqrt{D_1 D_2}}\esp{-\ui\frac{A_1^2}{D_2}+\ui\frac{A_2^2}{D_1}}
  = \frac1{2\ui\sqrt{D_1 D_2}}
  \esp{-\ui\frac{(A+A^*)^2}{4D_2}-\ui\frac{(A-A^*)^2}{4D_1}} \nonumber \\
  &= \frac{\esp{\ui A^+ A^-}}{2\ui\sqrt{D_1 D_2}} \;, \qquad
  A^\pm \equiv \frac{A_2}{\sqrt{D_1}} \pm \frac{A_1}{\sqrt{D_2}} \;.
  \label{Idifeq}
\end{align}
The integral can be reconstructed if we knew the boundary conditions.

An alternative method is to perform one of the $x$, $y$ integrals in~\eqref{igD}
by noticing that the exponent is bilinear in
$\xi,\eta\equiv\sqrt{D_2}x\mp\sqrt{D_1}y$, so that one variable $\xi$ or $\eta$
can be kept fixed and real, while the other is complexified and deformed on the
pole. By using
\begin{align*}
  &x=\frac{\xi+\eta}{2\sqrt{D_2}}\;, \quad y=\frac{\eta-\xi}{2\sqrt{D_1}}\;,
  \quad x-\ui y=\frac{\xi d +\eta d^*}{2\sqrt{D_1 D_2}} \;, \quad
  d\equiv \sqrt{D_1}+\ui\sqrt{D_2} =\sqrt{D_1+D_2}\esp{\ui\chi}\;,\\
  &2\zt\cdot\At = A_1\frac{\xi+\eta}{\sqrt{D_2}}+A_2\frac{\eta-\xi}{\sqrt{D_1}}
  = \xi\left(\frac{A_1}{\sqrt{D_2}}-\frac{A_2}{\sqrt{D_1}}\right)
  +\eta\left(\frac{A_1}{\sqrt{D_2}}+\frac{A_2}{\sqrt{D_1}}\right)
\end{align*}
and integrating over $\eta$ at fixed real $\xi$ at the double pole
$\eta=\esp{-\ui2\gamma}\xi$, $\gamma=\pi/2-\chi$, we obtain (in the case $A^+>0$)
\begin{align}
  I&= \int\frac{\dif\xi\dif\eta}{4\pi\ui\sqrt{D_1 D_2}}\;
  \frac{\esp{\ui\eta\xi}-1}{(\xi d+\eta d^*)^2}4D_1 D_2
  \esp{\ui\eta\left(\frac{A_1}{\sqrt{D_2}}+\frac{A_2}{\sqrt{D_1}}\right)
  +\ui\xi\left(\frac{A_1}{\sqrt{D_2}}-\frac{A_2}{\sqrt{D_1}}\right)} \nonumber\\
  &=\int_{-\infty}^0\dif\xi\;\frac{2\ui\sqrt{D_1 D_2}}{d^*{}^2}\left[
    \esp{\ui\xi\esp{-\ui2\gamma}(\xi+A^+)-\ui\xi A^-}(\xi+A^+)\Theta(\xi+A^+)
    -\esp{\ui\xi(\esp{-\ui2\gamma}A^+-A^-)} A^+\right] \nonumber \\
  &= \frac{2\sqrt{D_1 D_2}}{D_1+D_2}\left[
    \frac{\esp{-\ui2\gamma} A^+}{\esp{-\ui2\gamma}A^+ - A^-}\,
  -\ui\esp{-\ui2\gamma}A^+{}^2\int_0^1\dif\rho\;
  (1-\rho)\esp{-\ui\esp{-\ui2\gamma}A^+{}^2\rho(1-\rho)+\ui\rho A^+ A^-}\right]\;.
\end{align}

By performing a partial integration of the second term (so as to subtract the
term linear in $\rho$ of the integrand) we finally obtain
\begin{align}
  I &= \frac{2\sqrt{D_1 D_2}}{D_1+D_2}\left[\frac12\left(\esp{\ui A^+ A^-}
      -\frac{A^- + A^+\esp{-\ui2\gamma}}{A^- - A^+\esp{-\ui2\gamma}}
    \right)\right.\nonumber\\
    &\qquad\qquad\qquad\left. 
      -\frac{\ui}2 A^+(A^+\esp{-\ui2\gamma} + A^-)\int_0^1\dif\rho\;
      \esp{-\ui[A^+{}^2\esp{-\ui2\gamma}\rho(1-\rho)-A^+ A^-\rho]}\right]
    \nonumber\\
  &= \frac{C}{2}\left[ \esp{-\frac{\ui}{2}(\Ac^2+\Acb^2)}-\ui\frac{\Ac}{\Acb}
    +\ui\Ac\int_{-\ui\Acb}^\Ac\dif\Ac'\esp{-\frac{\ui}{2}(\Acb^2+\Ac'{}^2)}
    \right]
\end{align}
with the following variables:
\begin{align}
 \Ac&\equiv \frac{\ui}{\sqrt{2}}(A^+\esp{-\ui\gamma}+A^-\esp{\ui\gamma})
 = \sqrt{\frac{D_1+D_2}{2D_1 D_2}}\left(A + A^*\frac{D_1-D_2}{D_1+D_2}
 \right) \;, \label{Acdef}\\
 \Acb &\equiv \frac1{\sqrt{2}}(A^+\esp{-\ui\gamma}-A^-\esp{\ui\gamma})
 = \sqrt{\frac{2}{D_1+D_2}}A^* \\
 \Ac' &\equiv -\ui\Acb+ \ui\sqrt{2} A^+ \esp{-\ui\gamma}\rho
 \;, \qquad  C =\sin(2\gamma)=\frac{2\sqrt{D_1 D_2}}{D_1+D_2} \;.
\end{align}
Finally, an integration by parts shows that $I$ is identically given by
\begin{equation}\label{IAA}
  I = -C\frac{\Ac}{2}\int_{-\ui\Acb}^\Ac\frac{\dif \Ac'}{\Ac'{}^2}\;
  \esp{-\frac{\ui}{2}(\Acb^2-\Ac'{}^2)} \;.
\end{equation}
\begin{itemize}
\item It is easily verified that eq.~\eqref{IAA}, with the
  identification~\eqref{Acdef}, is the solution of the differential
  equation~\eqref{Idifeq} with boundary condition $\Ac_\ini=-\ui\Acb$.
\item If $D_1=1=D_2$, as in the case of the emission amplitude discussed in
  sec.~\ref{s:lorea}, $C=1$, $\Ac=A$ and $\Acb=A^*$. In particular,
  eq.~\eqref{IAA} reduces to eq.~\eqref{zetaInt}.
\item $\Ac$ and $\Acb$ are not complex conjugate for $D_1\neq D_2$.
\item The amplitude vanishes when the integration limits coincide, i.e.,
  $\Ac=-\ui\Acb$, corresponding to $A^+=A_2/\sqrt{D_1}+A_1/\sqrt{D_2}=0$, or
  equivalently $\phi_A = -\gamma$.  In the limit $D_1=D_2$, such nodal line
  corresponds to the azimuthal direction $\phi_A = -\pi/4$, while $\gamma$
  becomes possibly small for $D_1\ll D_2$, as depicted in
  fig.~\ref{f:azimQuadratic}.
\end{itemize}

\bibliographystyle{h-physrev5}
\bibliography{softBased}% Produces the bibliography via BibTeX.

\end{document}